\documentclass[journal]{IEEEtran}

\usepackage{xcolor,soul,framed} 

\colorlet{shadecolor}{yellow}
\usepackage[pdftex]{graphicx}
\graphicspath{{../pdf/}{../jpeg/}}
\DeclareGraphicsExtensions{.pdf,.jpeg,.png}

\usepackage[cmex10]{amsmath}
\usepackage{amsmath,amsfonts}
\usepackage{array}
\usepackage[caption=false,font=normalsize,labelfont=sf,textfont=sf]{subfig}
\usepackage{textcomp}
\usepackage{stfloats}
\usepackage{url}
\usepackage{verbatim}
\usepackage{graphicx}
\usepackage{cite}
\usepackage{adjustbox}
\usepackage{amssymb} 
\usepackage{multirow} 
\usepackage{bbm}
\usepackage{xcolor}

\usepackage{fancyhdr,graphicx,amsmath,amssymb}
\usepackage[ruled,vlined]{algorithm2e}
\usepackage{algpseudocode}

\begin{document}
\bstctlcite{IEEEexample:BSTcontrol}
    \title{Self-Supervised Human Activity Recognition with Localized Time-Frequency Contrastive Representation Learning}
  \author{Setareh Rahimi Taghanaki, Michael Rainbow, Ali Etemad

  \thanks{
  This work was partially funded by Natural Sciences and Engineering Research Council of Canada (NSERC).}
  \thanks{S. R. Taghanaki and A. Etemad are with the Department of Electrical and Computer Engineering as well as the Ingenuity Labs Research Institute, Queen's University, Kingston, Canada, {\tt\small \{19srt3, ali.etemad\}@queensu.ca}}%
  \thanks{M. Rainbow is with the Department of Mechanical Engineering, Queen's University, Kingston, Canada,  {\tt\small michael.rainbow@queensu.ca}}
}

\maketitle

\begin{abstract}
In this paper, we propose a self-supervised learning solution for human activity recognition with smartphone accelerometer data. We aim to develop a model that learns strong representations from accelerometer signals, in order to perform robust human activity classification, while reducing the model's reliance on class labels. Specifically, we intend to enable cross-dataset transfer learning such that our network pre-trained on a particular dataset can perform effective activity classification on other datasets (successive to a small amount of fine-tuning). To tackle this problem, we design our solution with the intention of learning as much information from the accelerometer signals as possible. As a result, we design two separate pipelines, one that learns the data in time-frequency domain, and the other in time-domain alone. In order to address the issues mentioned above in regards to cross-dataset transfer learning, we use self-supervised contrastive learning to train each of these streams. Next, each stream is fine-tuned for final classification, and eventually the two are fused to provide the final results. We evaluate the performance of the proposed solution on three datasets, namely MotionSense, HAPT, and HHAR, and demonstrate that our solution outperforms prior works in this field. We further evaluate the performance of the method in learning generalized features, by using MobiAct dataset for pre-training and the remaining three datasets for the downstream classification task, and show that the proposed solution achieves better performance in comparison with other self-supervised methods in cross-dataset transfer learning.

\end{abstract}

\begin{IEEEkeywords}
Self-supervised learning, contrastive learning, human activity recognition, accelerometers.
\end{IEEEkeywords}

\IEEEpeerreviewmaketitle

\section{Introduction}

\IEEEPARstart{H}{uman} activity recognition (HAR) has become a popular field of research due to its applicability in a number of settings such as smart homes \cite{fang2014recognizing, mehr2016resident}, wellness monitoring \cite{villeneuve2017reconstruction}, medical rehabilitation \cite{schrader2020advanced}, assisted living \cite{zhu2009multi}, and sports \cite{taghanaki2021wearable, hendry2020development}. While activity recognition can be carried out using data from a variety of sources such as videos or motion capture data, the pervasiveness of wearable devices such as smartphones, smartwatches, and other portable devices has created new research opportunities in HAR studies \cite{kulchyk2019activity, bock2021improving, saeed2019multi}.

Recent advancements in deep learning have inspired a variety of different solutions for HAR with most of them relying on fully supervised approaches \cite{wang2016comparative, ignatov2018real}. These methods, however, are highly reliant on large amounts of \textit{labeled} data for successful training. Yet, the process of labeling wearable human-centric data is generally time consuming and expensive. Furthermore, the characteristics of wearable signals make them difficult to interpret, even by human experts, in comparison to other modalities of data such as videos or audio, further complicating the notion of data annotation in this field.

The limitations stemming from lack of large annotated datasets has inspired researchers to develop alternative solutions that can alleviate the reliance on labeled data. Self-supervised learning \cite{saeed2019multi, haresamudram2020masked, rahimi2021self} is one of these techniques which has been proposed to tackle the problem by exploiting auxiliary tasks for pre-training, also referred to as pretext tasks, relying only on the \textit{unlabeled} data. In such approaches, in the first stage of training the neural network learns through pretext tasks, allowing it to extract general-purpose representations that can then be used in other related and similar tasks. The trained encoder is then frozen and transferred to be utilized as a feature extractor module in downstream tasks such as HAR. In this stage, a few dense layers are often added to the encoder to form a complete classifier or regressor for the downstream task. However, in contrast to the fully-supervised counterpart networks, considerably smaller amounts of labeled data are required to train the dense layers as the number of trainable parameters in these layers is considerably less than the entire network. A recently proposed approach to self-supervised learning named contrastive learning \cite{chen2020simple, grill2020bootstrap, chen2021exploring} relies on two different sets of views crafted from the input instance using \textit{augmentations}. The goal then is to train the network such that it could distinguish between the views generated from the \textit{same} input instance versus the ones from \textit{different} input instances. This approach allows the model to learn to extract generalized representations which could be used for downstream tasks.

\vspace{5pt}
\noindent \textbf{Challenges.}
Self-supervised frameworks have been recently used in a few prior works for HAR with accelerometer signals \cite{khaertdinov2021contrastive, liu2021contrastive, saeed2020federated}. However, a number of open challenges and research questions motivate our work in this paper. First and foremost, there are a number of design choices and parameters which need to be carefully crafted in self-supervised learning, for instance the use of negative pairs, augmentations, and others. Second, it is well-documented that when it comes to time-series (e.g., accelerometer signals) important information can be found in both time and frequency domains. In this context, it is important to explore how contrastive learning can learn representations in both domains, and how they can be aggregated to yield strong downstream results. 
Lastly, the notion of knowledge generalization or transfer across datasets is generally a very important yet challenging problem when it comes to human-generated data such as accelerometer signals, which we intend to achieve in this work.

\vspace{5pt}
\noindent \textbf{Contributions.} 
In this study we present a new self-supervised solution for HAR based on contrastive learning. In order to learn strong representations in both time and frequency domains with \textit{temporal localization}, our method first transforms the multi-dimensional Inertial Measurement Units (IMU) time-series into scalograms using a Morlet wavelet. Two contrastive architectures, which we name the signal learner and the scalogram learner respectively, exploit a contrastive approach to learn the human activity space in each domain separately. In the scalogram learner, time-frequency augmentations are used to generate two different views from each scalogram in an input batch. Each of the two views are then fed to a 2-dimensional (2D) convolutional encoder followed by a projection head, resulting in two sets of representations in a lower dimensional latent space. The agreement between the two latent representations is then contrastively maximized by grouping similar embeddings together while pushing diverse ones far from each other. The signal learner has a similar structure but with a few modifications. As it takes the time-series signals as inputs, it applies a set of temporal augmentations (instead of time-frequency augmentation) and uses a 1-dimensional (1D) convolutional layers as its encoder. Once the the two contrastive pipelines are pre-trained in a self-supervised manner, the two encoders are frozen and transferred to the downstream stage as feature extractors where they are accompanied by new sets of fully connected (FC) layers. Each network is then slightly fine-tuned with labeled data to learn to classify different human activity classes, and the two pipelines are fused using later fusion. 

In summary, we make the following contributions: 
\begin{itemize}

\item We propose a new solution for HAR based on self-supervised learning. Our model learns to extract effective and general-purpose latent representations in both time and localized time-frequency domains. This is achieved by using two separate architectures, a \textit{signal learner} and a \textit{scalogram learner}, which are trained separately using contrastive learning. Downstream, successive to adding the activity classification multilayer perceptron (MLP) heads and fine-tuning, the outputs of the two streams are fused with a simple late-fusion strategy.

\item We evaluate our method on three publicly available human activity datasets, MotionSense \cite{malekzadeh2018protecting}, HAPT \cite{reyes2016transition}, and HHAR \cite{stisen2015smart}. We perform thorough analysis of our solution on these datasets and compare our work to other self-supervised methods. Our approach outperforms the other methods on all three datasets, setting new \textit{state-of-the-art}. We perform detailed ablation studies to study the impact of each component of our network and observe that both the signal and scalogram learners play key roles in the better performance of our method.

\item Furthermore, we evaluate the robustness and generalization ability of the learnt representations by transferring the knowledge across datasets. To do so, we train our solution on a separate dataset, MobiAct \cite{chatzaki2016human}, and then evaluate the method on MotionSense , HAPT , and HHAR . Our results show improvement with respect to other methods in the area, demonstrating the generalization of our method.

\end{itemize}

The remainder of the paper is organized as follows. In Section \ref{Related_work} an overview of the previous works in HAR, self-supervised learning, and contrastive learning is presented. A detailed description of our proposed solution is provided in Section \ref{journal_method_section}. This is followed by the experiments and discussion on the obtained results in Section \ref{Experiment}. Lastly Section \ref{Conclusion} presents a summery of our work and potential future directions of inquiry.

\section{Related Work}\label{Related_work}

In this paper, we present a review of HAR with deep learning methods. First we present a short summary of prior work that have used fully-supervised solutions to train the solutions. This is followed by a general description of self-supervised learning, and subsequently the used of self-supervised learning for HAR.

\subsection{Fully-supervised HAR}

The success of deep neural networks in areas such as computer vision \cite{xu2015discriminative, lee2017going} and natural language processing \cite{collobert2011natural, huang2015bidirectional} has encouraged the adoption of such methods for human activity recognition using wearables or smartphone IMU data \cite{zhang2022deep, xia2020lstm, bianchi2019iot}. For instance,
convolutional neural networks (CNN) \cite{bianchi2019iot, yao2018efficient, kulchyk2019activity} and recurrent neural networks (RNN) \cite{xia2020lstm, mekruksavanich2020smartwatch} have been widely used for wearable HAR, and achieved strong performances in learning discriminative representations and subsequent classification of activities.

In \cite{bianchi2019iot}, 1D convolutional layers were used to perform activity classification with the aim of personalized HAR. Their method was effective in performing HAR robustly in the context of smart homes. In \cite{yao2018efficient}, variable-sized windows were used with a 5-layer CNN. Their method was able to overcome the challenges raised by fix-sized windows as well as certain windows containing data from multiple classes. In \cite{kulchyk2019activity}, an IMU-based activity recognition classification framework was designed using a simple 5-layer CNN. Furthermore, the number of IMUs and their placement on different parts of the body was studied, identifying that a minimum of two sensors with at least one of them located on body parts with wide range of motions (e.g. ankle) is required to achieve robust classification. A CNN followed by an RNN was presented in \cite{xia2020lstm} where 2 convolutional layers and 2 long short-term memory (LSTM) layers were combined to implement a model for HAR with smartphone data. The temporal characteristics of LSTM alongside the replacement of an FC layer with a global average pooling layer resulted in high accuracy of the model with a relatively small parameter set. Similarly, a spatio-temporal model was developed in \cite{mekruksavanich2020smartwatch}. In this method, two 1D convolutional layers were used followed by an LSTM layer, and dropout has been used to prevent overfitting. A comprehensive review of supervised HAR is available in \cite{zhang2022deep}.

\begin{figure}[t!]
\centerline{\includegraphics[width=1\columnwidth]{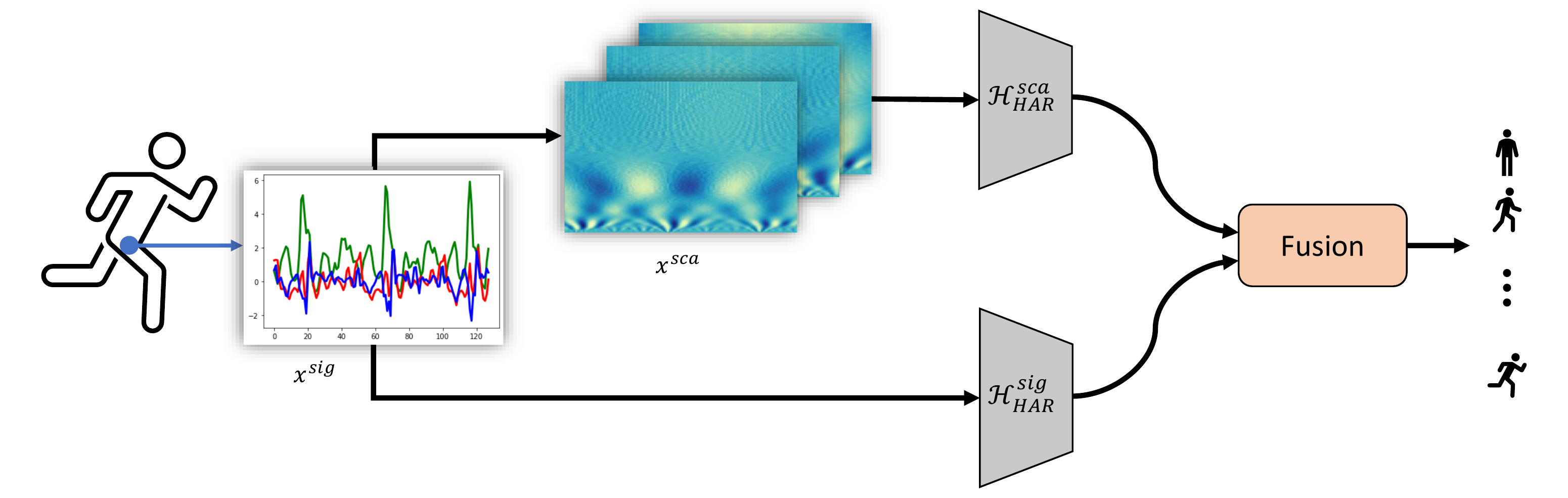}}
\caption{\textcolor{black}{A high-level architecture of our human activity recognition model ($\mathcal{F}$) is presented. Activity classification is performed by fusing learned features from input accelerometer signals, extracted by $\mathcal{H}^{sca}_{HAR}$ and $\mathcal{H}^{sig}_{HAR}$.}} 
\label{fig:arch_highLevel}
\end{figure}

\subsection{Self-supervised HAR}

\noindent \textbf{General background.} 
While traditional fully-supervised models have shown remarkable performance for HAR, they heavily rely on large amounts of labeled training data. Since data annotation is usually an expensive and time consuming process, recent studies have been motivated to explore alternative approaches that can relieve the reliance on annotations. Self-supervised learning is an alternative method to fully-supervised training, that avoids the need for expensive annotations by utilizing large-scale unlabeled data to learn generalized representations \cite{jing2020self, jaiswal2020survey}.

Self-supervised learning is a general learning framework in which the underlying representations of unlabeled data are learned through surrogate tasks called \textit{pretext} \cite{jing2020self}. Pretext learning is the first stage of self-supervision, which is carried out to learn generalized representations using automatically generated pseudolabels. In this stage a classification or regression task is often designed using which the neural network is trained. Through learning to solve the pretext tasks, the model learns to extract representations that can be utilized in the next stage of the framework for other related tasks called \textit{downstream}. The trained model is then frozen and subsequently transferred to the downstream pipeline, where a few untrained dense layers are added, and the whole pipeline is fine-tuned to solve the desired downstream tasks. Examples of pretext tasks include masked reconstruction \cite{haresamudram2020masked}, puzzle solving \cite{noroozi2016unsupervised}, transformation classification \cite{gidaris2018unsupervised, saeed2019multi}, and others.

A different category of self-supervised learning approaches called contrastive learning approaches rely on Siamese architectures during pretext learning, where they aim to group similar samples together while placing diverse samples far from each other. Here, augmentations are first applied on input individual samples to generate augmented views. Views from the same sample then form a positive pair, while negative pairs consist of views generated from different samples. Encoder networks are then used to extract representations from a given pair, generating two embeddings. Lastly the model learns to discriminate the positive pairs from the negative ones based on measuring the similarity of the two generated embeddings \cite{jaiswal2020survey}. 
Examples of various contrastive approaches include SimCLR \cite{chen2020simple}, SimSiam \cite{chen2021exploring}, and others, each of which adapt a different loss and pairing strategy for self-supervision. 
These techniques are further explored and described in Section \ref{Experiment}.

\vspace{5pt}

\noindent \textbf{HAR using self-supervised learning.} 
While the majority of prior works on HAR have relied on fully-supervised methods, a few notable studies have performed wearable-based HAR using self-supervised learning in recent years \cite{saeed2019multi, haresamudram2020masked, rahimi2021self}. The idea of using self-supervised learning in this field was first explored in \cite{saeed2019multi}. A multi-task classification problem was designed as the pretext task in the self-supervised setup. Eight signal transformations were separately applied to accelerometer signals to generate the inputs of the pretext model. The pretext model was built using 1D convolution layers as the encoder and task-specified MLP layers for classification. The model was trained to learn whether any of the transformations were applied on the input signal or not. The encoder layers were then frozen, and a few fully connected layers were added. Their method was able to successfully learn generalized representations from unlabeled data and improve performance when few labeled samples were available.

A masked reconstruction method was proposed in \cite{haresamudram2020masked}. In this work, certain data-points of windowed accelerometer and gyroscope data were randomly masked and reconstructed as the pretext self-supervised task. Ten percent of the data-points in each window were set to zero and the pretext network was trained to rebuild the masked data-points, through which it also learned to extract underlying representations by capturing the local temporal dependencies in the signal window. In a similar approach, a cross-dimensional motion prediction pretext framework was proposed in \cite{rahimi2021self}. A deep convolutional neural network was trained to predict an entire segment of accelerometer time-series in the $z$ axis (consisting of several consecutive samples), given the data in $x$ and $y$ axes as well as prior values of $z$ as inputs. It was shown in both \cite{haresamudram2020masked} and \cite{rahimi2021self} that masked reconstruction is a viable approach for IMU-based HAR in self-supervised frameworks.

Lastly, a few recent works have relied on contrastive learning \cite{khaertdinov2021contrastive}, \cite{wang2021sensor}. In these methods, data augmentations have been applied to IMU data in the pretext stage, following downstream fine-tuning for HAR. Both solutions have relied on SimCLR as their main contrastive framework, and demonstrated promising results. It should be noted that while both \cite{khaertdinov2021contrastive} and \cite{wang2021sensor} have performed the pretext contrastive stage in time domain, an earlier work which had performed HAR using federated learning and with multimodal data (electroencephalography, blood volume pulse, and WiFi channel state alongside accelerometer), had suggested that exploiting frequency domain information in a contrastive setup can benefit activity recognition \cite{saeed2020federated}. Motivated by this notion, we propose the use of both time and frequency domain information for contrastive learning. As we will describe later, our use of frequency domain information is quite different than that of \cite{saeed2020federated}. While frequency domain projections are used as augmentations themselves in \cite{saeed2020federated}, we explicitly learn representations in both time and frequency domains in our framework following localized time-frequency transforms and augmentations in each domain. \textcolor{black}{Moreover, it should be noted that while \cite{khaertdinov2021contrastive} and \cite{wang2021sensor} only explore SimCLR and MOCO, we experiment with SimCLR and SimSiam, given the recent success of SimSiam in a variety of other domains.}

\section{Method\label{journal_method_section}}

\subsection{Solution Overview}
Our goal is to develop a model $\mathcal{F}$ for HAR that learns robust representations $e$ from accelerometer signals $x$, while alleviating the need for class labels. Furthermore, we aim to explore cross-dataset transfer learning, i.e., when $\mathcal{F}$ is pretrained on dataset $D_i$, it can perform robust HAR on dataset $D_j$, where $i \neq j$, while remaining mostly frozen (after a small degree of fine-tuning). Moreover, we hypothesize that important activity-related information reside in both time and frequency domains. As a result, we aim to effectively exploit both these domains for self-supervised learning. To this end, we implement $\mathcal{F}$ in a way that extracts efficient representations from input accelerometer signals in frequency domain with \textit{temporal localization} in addition to time domain, and then aggregates them to yield robust HAR performance. 
To achieve this, two contrastive architectures named scalogram learner and signal learner are implemented to extract motion-related representations in both domains. These networks are first pre-trained with unlabeled data to learn the contrastive pretext task, and then frozen and fine-tuned with labeled data to learn the HAR task. Lastly, score-level fusion is applied to make the final decision. In the next few subsections we describe the different components of our solution in detail. A high-level architecture of ($\mathcal{F}$) is presented in Figure \ref{fig:arch_highLevel}.


\subsection{Scalogram Learner}
When analyzing time-series, important information is generally manifested in both time and frequency domains. While Fourier Transform (FT) is one of the most widely used techniques in analyzing the frequency content of a signal, it does not retain important time information. To counter this and represent the signal in a joint time-frequency domain, techniques such as Short-Time-Fourier-Transform (STFT) and wavelet transform (which produces scalograms) can be used. Both methods retrieve the time-frequency content of the signal by dividing it into time windows and operating on each window separately. However, since the wavelet transform processes a signal at different frequencies with different resolutions, it is better at localizing time-frequency properties \cite{daubechies1990wavelet, merry2005wavelet, nedorubova2021human}.

\begin{figure}[t!]
	\centering
	\includegraphics[width=0.7\columnwidth]{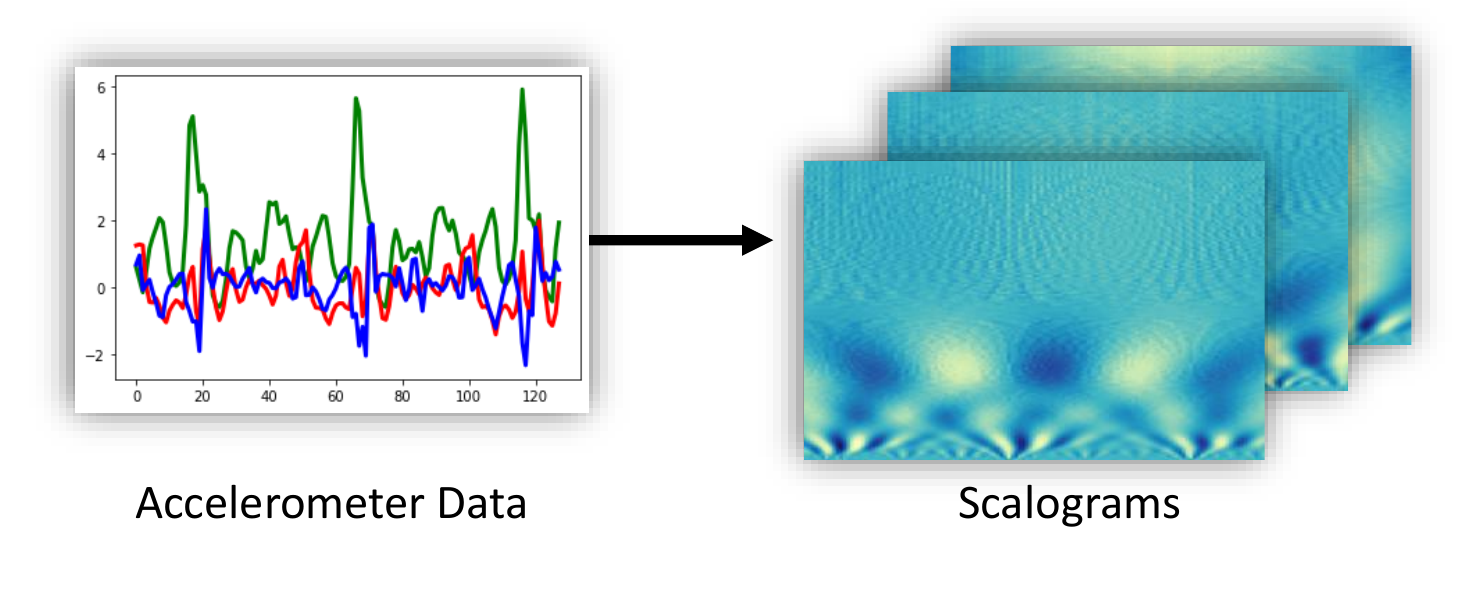}
	\caption{\textcolor{black}{Sample input tri-axis accelerometer signals and the corresponding scalograms generated using the Morlet wavelet transform.}}
	\label{signalToscalogram}
\end{figure}

Accordingly, we generate the scalogram of the input accelerometer signals by applying Continuous Wavelet Transform (CWT). We use Morlet transform as the wavelet function to decompose the input signal into a combinations of wavelets. The wavelet transform of signal $x(t)$ is given by:
\begin{equation}
\label{wavelet function}
x^{sca}(a,b) = \frac{1}{\sqrt{a}}\int_{-\infty }^{+\infty } x^{sig}.\psi (\frac{t-b}{a})dt, 
\end{equation}
where $\psi$ is the wavelet function, and $a$ and $b$ are the scaling and translation factors. The Morlet wavelet $\psi$ is defined as follows:
\begin{equation}
\label{Morlet_wavelet}
\psi  \left ( t \right ) = exp^{-\frac{t^{2}}{2}}cos(5t).
\end{equation}
\textcolor{black}{Figure \ref{signalToscalogram} illustrates a sample 3-channel scalogram generated from the 3 dimensions ($x$, $y$, $z$) of the raw accelerometer signals.}

Successive to generating scalograms of the accelerometer time-series, we design a model for learning effective scalogram representations, henceforth referred to as the \textit{scalogram learner}. In our solution, in order to exploit unlabeled data and reduce reliance on labeled samples, we use self-supervised contrastive learning. Thus, the scalogram learner is denoted by $\mathcal{H}^{sca}_{cont}$.

\begin{figure*}[t!]
\centerline{\includegraphics[width=0.8\linewidth]{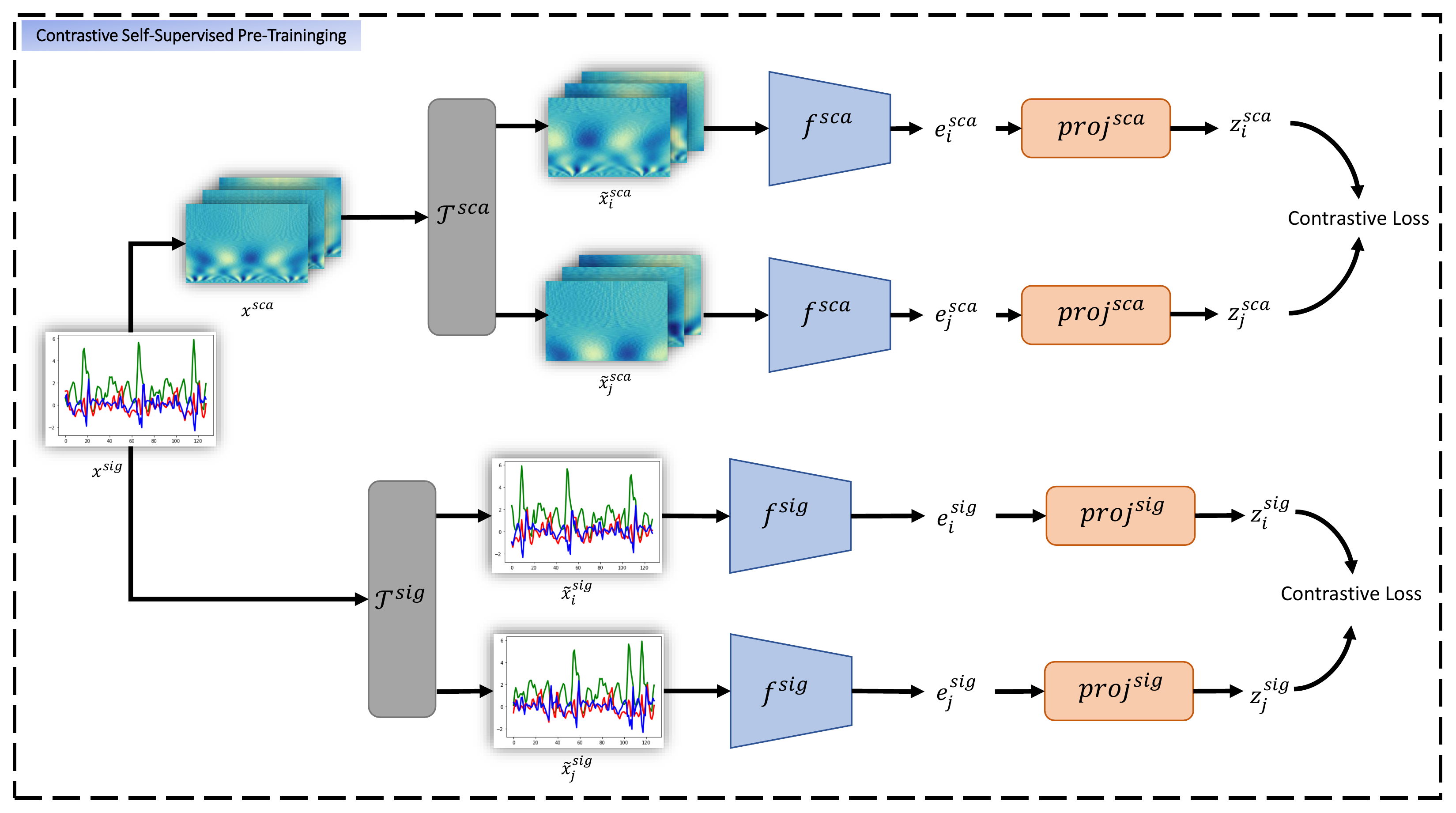}}
\caption{\textcolor{black}{The architecture of the self-supervised networks, signal learner ($\mathcal{H}^{sig}_{cont}$) and scalogram learner ($\mathcal{H}^{sca}_{cont}$), are presented. These two contrastive pipelines are separately pre-trained on unlabeled accelerometer signals and their corresponding scalograms. As a result, Two encoders, $f^{sig}$ and $f^{scal}$, are trained to learn effective representations in both time and time-frequency domains. The learned weights of the encoders are then frozen and transferred for the downstream task.}} 
\label{fig:journal_pretext}
\end{figure*}

We adopt the SimCLR framework \cite{chen2020simple} for our solution and design $\mathcal{H}^{sca}_{cont}$. In this framework, the generated scalograms are first augmented in the time-frequency domain. Next, an encoder is used to learn representations from the augmented samples. These representations are then mapped onto a latent space, whose loss is minimized using a contrastive loss function. This component of our proposed solution is presented in Figure \ref{fig:journal_pretext}(a).
Below we describe each of these components in detail.


\vspace{5pt}
\noindent \textbf{Time-Frequency Augmentation.}
Given that the scalogram $x^{sca}$ of an input signal is represented as a 2D matrix, it can be depicted as an image. As a result, augmentations that have been widely employed on images for contrastive learning can be applied in this context. However, it should be noted that by utilizing such augmentations, we are effectively applying time-frequency augmentations to the scalograms. This augmentation module, $\mathcal{T}^{sca}$, generates two views of $x^{sca}$, denoted by $\tilde{x}^{sca}_{i}$ and $\tilde{x}^{sca}_{j}$. The transformations used in this study (motivated by \cite{chen2020simple}) are described bellow:

\begin{itemize}
\item{\textit{Random color distortions:}  This transformation changes the appearance of the scalograms and is composed by color dropping (converting the scalograms from RGB to grayscale) and color jittering (adjusting the brightness, contrast, saturation, and hue of the scalograms by \textcolor{black}{random factors in the ranges of (-0.9, 0.9), (0.1, 1.9), (0.1, 1.9), and (-0.3, 0.3), respectively}).}
\item{\textit{Random cropping and resizing:} This geometric transformation crops the scalograms randomly and resizes them to their original resolution.}
\item{\textit{Flip:} This transformation flips the scalograms horizontally.}
\end{itemize}

\vspace{5pt}
\noindent \textbf{Scalogram Encoder.}
Assuming $\tilde{x}^{sca}_{i}$ and $\tilde{x}^{sca}_{j}$ being 2 augmented data samples, a 2D convolutional neural network encoder, $f^{scal}$, takes them as inputs and extracts representation vectors $e^{sca}_i = f^{scal}(\tilde{x}^{sca}_{i})$ and $e^{sca}_j = f^{scal}(\tilde{x}^{sca}_{j})$. The implementation details of the network are presented in Table \ref{tab:SSL_arch_2d}. The scalogram encoder $f^{scal}$ contains three 2D convolutional layers with kernel sizes set to 8, 4, and 4, and the number of filters set to 32, 64, and 96, followed by a 2D global maxpooling layer. L2 regularization with a rate of 0.0001 is applied to the convolutional layers.

\begin{table}[t]
    \centering
    \caption{\textcolor{black}{Architecture and parameters of the scalogram encoder ($f^{sca}$) and projection head ($proj^{sca}$), along with the HAR MLP head are presented.}}
    \begin{tabular}{l|l|l}
    \hline
     \textbf{Module} & \textbf{Layer Details} & \textbf{Feature Shape} \\ \hline \hline
      Input & $-$ & $128 \times 128 \times 3$ \\ \hline
      \multirow{4}{*}{Encoder} & $[\textit{conv}, 8 \times 8, 32]$ &   $121 \times 121 \times 32$  \\ 
                                     &  $[\textit{conv}, 4 \times 4, 64]$ &   $118 \times 118 \times 64$  \\ 
                                     &  $[\textit{conv}, 4 \times 4, 96]$ &   $115 \times 115 \times 96$ \\
                                     & \textit{global maxpool} &  $96$ 
                                     \\ \hline 
     \multirow{3}{*}{Projection Head} &  $[\textit{dense}, 96]$ &   $96$  \\ 
                                            &  $[\textit{dense}, 96]$ &   $96$  
 \\ \hline 
     \multirow{5}{*}{HAR} &  $[\textit{dense}, 256 ] $ &   $256$  \\ 
                                            &  $[\textit{dense}, num\_classes]$ &   $num\_classes$  \\
                                            & \textit{output} &  1
 \\ \hline 

    \end{tabular}
    \label{tab:SSL_arch_2d}
\end{table}

\vspace{5pt}
\noindent \textbf{Scalogram Projection Head.}
The next step in our contrastive framework is mapping the representations to a latent space for the contrastive loss to then be applied. In this step, an MLP-based projection head module, $proj^{scal}$, containing two 96-neuron FC layers is applied on the representations, mapping them onto a space ${z^{sca}_i} = proj^{sca}(e^{sca}_i)$ and ${z^{sca}_j} = proj^{sca}(e^{sca}_j)$.

\vspace{5pt}
\noindent \textbf{Loss Function.}
Lastly, a loss function is used to group the differently augmented views of the same data example and push away the ones from different data samples in the latent space. Our contrastive framework is trained with the Normalized Temperature-scaled Cross Entropy \textcolor{black}{\cite{chen2020simple, chen2017sampling}}, 
which for a positive pair of examples ${(i, j)}$ is given by:
\begin{equation}
\label{Loss_function}
\ell(i,j) = -log\frac{exp(sim(z^{sca}_i,z^{sca}_j)/\tau)}{\sum_{k=1}^{2N} \mathbbm{1}_{[k\neq i]} exp(sim(z^{sca}_j,z^{sca}_k)/\tau)},
\end{equation}
where the cosine similarity function $sim(z_{i},z_{j})$ is given by:
\begin{equation}
\label{similarity function}
sim(z^{sca}_i,z^{sca}_j) = (z^{sca}_i)^{T}z^{sca}_j/\left\| z^{sca}_i \right\|\left\| z^{sca}_j \right\|.
\end{equation}
Accordingly, the total loss is defined by:
\begin{equation}
\mathcal{L} = \frac{1}{2N}\sum_{k=1}^{N}[\ell(2k-1,2k) + \ell(2k,2k-1)],
\end{equation}
where $N$ is the batch size. 

\subsection{Signal Learner}
As discussed earlier, our solution consists of two learner for learning both time and localized time-frequency information. Therefore, we implement a second contrastive model called the \textit{signal learner}, $\mathcal{H}^{sig}_{cont}$, with the purpose of learning representations solely in time domain. Figure \ref{fig:journal_pretext}(b) depicts this component of our solution.
Similar to the scalogram learner, $\mathcal{H}^{sig}_{cont}$ comprises of four components:

\vspace{5pt}
\noindent \textbf{Temporal Augmentations.} Given that the input accelerometer signals are time-series, temporal augmentation can be applied \cite{tang2020exploring}. Thus, the time augmentation module $\mathcal{T}^{sig}$ is applied on an input signal $x^{sig}$ to generate two views of the signal $\tilde{x}^{sig}_i$ and $\tilde{x}^{sig}_j$. The time transformations used in this study (motivated by \cite{tang2020exploring}) are described as follows:

\begin{itemize}
\item{\textit{Noise:} This transformation adds random Gaussian noises with a mean of 0 and standard deviation of 0.1 to the original signal.}
\item{\textit{Scale:} This transformation multiplies a random factor as scaler with a mean of 1 and standard deviation of 0.2 by the magnitude of the samples.}
\item{\textit{Negation:} This transformation scales the original signals by a factor -1.}
\item{\textit{Time-Flip:} This transformation reverses the direction of the signal by reversing the samples within the sample along the time axis.}
\item{\textit{Channel-Shuffle:} This transformation randomly shuffles the channels of the signal.}
\item{\textit{Permutation:} This transformation slices the signal window samples into different samples and then randomly reorders them to generate a new signal.}
\item{\textit{Rotation:} This transformation rotates the signal by a random degree around a random axis in 3D space.}  
\item{\textit{Time-Warp:} This transformation locally stretches or warps a time-series through a smooth distortion of time intervals between the values.}
\end{itemize}

\vspace{5pt}
\noindent \textbf{Signal Encoder}
The augmented signal samples $\tilde{x}^{sig}_{i}$ and $\tilde{x}^{sig}_{j}$ are subsequently fed to a 1D convolution-based neural network encoder $f^{sig}$, yielding representation vectors ${e^{sig}_i} = f^{sig}(\tilde{x}^{sig}_i)$ and ${e^{sig}_j} = f^{sig}(\tilde{x}^{sig}_{j})$. As presented in Table \ref{tab:SSL_arch_1d}, $f^{sig}$ is built using three convolutional layers with kernel sizes set to 12, 8, and 8, and number of filters set to 32, 64, and 96, followed by a 1D global maxpooling layer. L2 regularization with a rate of 0.0001 is applied to the convolutional layers.


\begin{figure*}[t!]
\centerline{\includegraphics[width=0.7\linewidth]{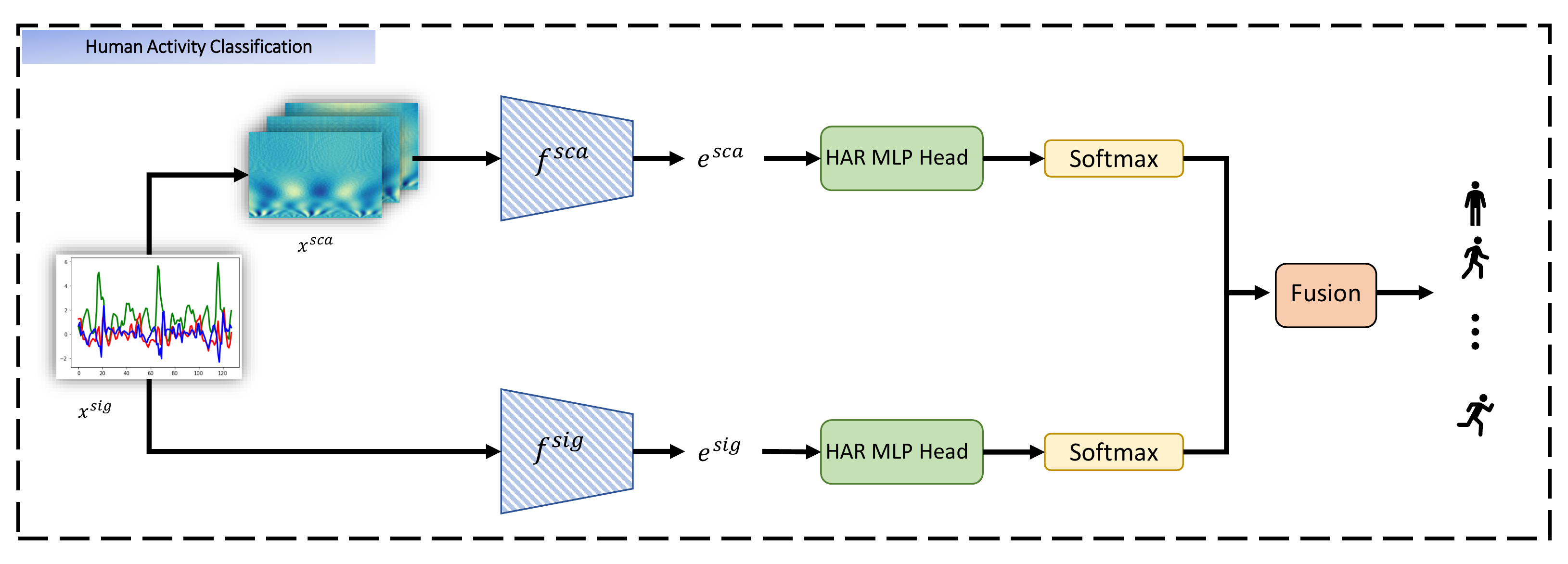}}
\caption{\textcolor{black}{Detailed architecture of the downstream human activity recognition model ($\mathcal{F}$) is presented. The encoders $f^{sig}$ and $f^{sca}$ have been pre-trained in the pretext stage, and their frozen layers are utilized as feature extractors for the input signal $x^{sig}$ and the corresponding scalogram $x^{sca}$.}}
\label{fig:arch_signal}
\end{figure*}

\begin{table}[t]
    \centering
    \caption{\textcolor{black}{Architecture and parameters of the signal encoder ($f^{sig}$) and projection head ($proj^{sig}$), along with the HAR MLP head are presented.}}
    \begin{tabular}{l|l|l}
    \hline
     \textbf{Module} & \textbf{Layer Details} & \textbf{Feature Shape} \\ \hline \hline
      Input & $-$ & $128 \times 3$ \\ \hline
      \multirow{4}{*}{Encoder} & $[\textit{conv}, 1 \times 12, 32]$ &   $116 \times 128$  \\ 
                                     &  $[\textit{conv}, 1 \times 8, 64]$ &   $54 \times 256$  \\ 
                                     &  $[\textit{conv}, 1 \times 8, 96]$ &   $23 \times 384$ \\
                                     & \textit{global maxpool} &  $96$ 
                                     \\ \hline 
     \multirow{3}{*}{Projection Head} &  $[\textit{dense}, 96]$ &   $96$  \\ 
                                            &  $[\textit{dense}, 96]$ &   $96$  
 \\ \hline 
     \multirow{5}{*}{HAR} &  $[\textit{dense}, 256 ] $ &   $256$  \\ 
                                            &  $[\textit{dense}, num\_classes]$ &   $num\_classes$  \\
                                            & \textit{output} &  1
 \\ \hline 

    \end{tabular}
    \label{tab:SSL_arch_1d}
\end{table}

\vspace{5pt}
\noindent \textbf{Signal Projection Head}
In this step, an MLP-based projection head module, $proj^{sig}$, which contains two 96-neuron FC layers is used to map the representations onto a space ${z^{sig}_i} = proj^{sig}(e^{sig}_i)$ and ${z^{sig}_j} = proj^{sig}(e^{sig}_j)$.

\vspace{5pt}
\noindent \textbf{Loss Function}
Similar to the scalogram learner, simCLR loss \ref{Loss_function} is used to on the representations with the objective of grouping augmented views of the same sample and pushing away the ones from different samples. The difference with respect to Equation \ref{Loss_function} is that instead of $z^{sca}_i,z^{sca}_j$, we use $z^{sig}_i,z^{sig}_j$ for this model.


\subsection{Downstream HAR and Fusion}
After training both the scalogram and signal learners using contrastive learning in a self-supervised manner, we freeze the weights of the two encoders $f^{sca}$ and $f^{sig}$, and discard the rest of the pre-text pipeline including $proj^{sca}$ and $proj^{sig}$. Next we add a HAR MLP head to $f^{sca}$, which consists of two FC layers, one with 256 units and the other with the size of the number of activity classes. Finally a softmax layer is added to the network. This overall pipeline is henceforth referred to $\mathcal{H}^{sca}_{HAR}$. In the same manner, we construct $\mathcal{H}^{sig}_{HAR}$ by taking the frozen $f^{sig}$ and adding a HAR MLP head, which consists of two FC layers (similar in size to the scalogram HAR MLP head), followed by a softmax layer. These two downstream HAR networks are depicted in Figure \ref{fig:arch_signal}.

Next, we fine-tune each pipeline by supervised training the two HAR MLP heads with class information. Our experiments later show that when fine-tuning, the model benefits from a slight defrosting of the final convolution layer and allowing it to also re-train along with the HAR MLP heads to allow for better compatibility. As we will show later in Section \ref{Experiment}, this fine-tuning can be done on the same dataset used to pre-train the scalogram and signal learners in self-supervised manner, or even with completely new datasets, enabling cross-dataset transfer learning with strong performances. This training step is done using categorical cross-entropy loss.
Finally, to combine the two pipelines for activity classification, we perform fusion between the two streams. As we will illustrate later in Section \ref{Experiment}, both early (feature-level) and late (score-level) fusions were considered. Our experiments, however, show that score-level fusion provides more reliable results, which we use in our final solution (see Figure \ref{fig:arch_signal}).

\subsection{Training Details}
We implement our model using Keras with TensorFlow backend and perform the training on NVIDIA GeForce RTX 2080 Ti GPUs. We use a batch size of 128 and Adam optimizer with a learning rate $= 0.001$ to train $\mathcal{H}^{sca}_{cont}$, $\mathcal{H}^{sig}_{cont}$, $\mathcal{H}^{sca}_{HAR}$, and $\mathcal{H}^{sig}_{HAR}$, for 50, 150, 50, and 70 epochs, respectively. Furthermore, during training of HAR models, early-stopping is applied to avoid overfitting. Rectified Linear Unit (ReLU) activation is applied on all the layers, except for the output layers. Training hyperparameters are tuned empirically to obtain optimum performance.

\subsection{Validation Strategy}
To train and evaluate our solution on different datasets, we use two specific validation strategies, henceforth referred to as Scheme 1 and Scheme 2.
Both strategies are user-independent, in that data from a particular user cannot appear in both training and testing. 
\begin{itemize}
\item In \textbf{\textit{Scheme 1}}, 
we employ a user-split based 5-fold cross validation (\textit{leave-some-subjects-out}), where 
20\% of the subjects in each dataset are randomly selected and used in the testing phase, while the remaining 80\% are utilized for training. Each experiment is repeated 5 times with different users in the testing split at each run. This approach has been used in prior works such as \cite{saeed2019multi}, \cite{haresamudram2021contrastive}, and \cite{rahimi2021self}. Moreover, in the training phase, the training set is split into the train and the validation sets with 20\% of the data allocated to the validation split. In the end, we report the average of the 5 iterations. 
\item In \textbf{\textit{Scheme 2}}, we randomly select approximately 20\% of the users as `test-users' and train our solution on the remaining users. While we acknowledge that this scheme bears a resemblance to Scheme 1, this approach has been widely used in the literature, by prior works such as \cite{saeed2019multi}, \cite{haresamudram2020masked}, and \cite{tang2020exploring}. Therefore, in order to provide a fair comparison, we also used this strategy.
\end{itemize}

We calculate two metrics for evaluation purposes, standard weighted F1 and Cohen’s kappa ($K$), which is calculated by $K = Acc_{T} - Acc_{R}/1-Acc_{R}$, where $Acc_{T}$ is the total accuracy of the model, indicating the probability of agreement between the model predictions and the ground truth values, and $Acc_{R}$ denoting random accuracy, indicating the probability of both predictions and the ground truth values agreeing on the labels by chance.

\section{Experiment and Results}\label{Experiment}

In this section we describe the details of the experiments for pre-training and downstream activity classification, along with results on various datasets and comparisons to prior work in this area. Additionally we perform a number of experiments such as ablations, and additionally test the sensitivity of our solution with respect to the amount of labeled samples used for training.

\subsection{Datasets}
We evaluate our method on four publicly available human activity datasets, MotionSense, HAPT, HHAR, and MobiAct. The details of each dataset including activity classes, number of subjects, and the electronic devices used to collect the data are briefly described in the following.


\subsubsection{\textbf{MotionSense}}
This dataset consists of time-series data generated by accelerometer and gyroscope sensors collected with an iPhone 6s using SensingKit. 6 daily activities (walking downstairs, walking upstairs, walking, sitting, standing, and jogging) were performed by 24 participants in 15 trials under similar environments and conditions while the smartphone was kept in the participants' front pocket to record the motion data.

\subsubsection{\textbf{HAPT}}
The Human Activities and Postural Transitions (HAPT) dataset comprises accelerometer and gyroscope signals of 12 activities (6 basic activities: walking, walking downstairs, walking upstairs, standing, sitting, and lying, and 6 postural transitions: stand-to-sit, sit-to-stand, sit-to-lie, lie-to-sit, stand-to-lie, and lie-to-stand). The data was collected from 30 participants with waist-worn Samsung Galaxy S2 devices.

\subsubsection{\textbf{HHAR}}
This dataset consists of accelerometer and gyroscope generated data collected with 8 smartphones and 4 smartwatches worn by 9 participants. The participants performed 6 activities (sitting, standing, walking, stairs-up, stairs-down, and biking) each for 5 minutes while carrying the phones around their waist and wearing 2 watches on each arm. In this study, we only used the data collected by the smartphones.

\subsubsection{\textbf{MobiAct}}
This dataset consists of accelerometer and gyroscope signals of 11 daily activities (walking, jogging, jumping, upstairs, downstairs, sitting, stand to sit, sit to stand, sitting on a chair, car step-in, and car step-out) and 4 types of falls collected using Samsung Galaxy S3 phones through more than 3200 trials. 66 participants of varying ages, genders, and weights were asked to carry the smartphone in their pockets to mimic the everyday usage of the device. In this study we only utilize the data of the subjects who have samples from all of the 11 activities.

\subsection{Performance and Comparison}
In this subsection we present the results of our experiments. 
First, we compare our work to other methods in the area. Next, we study our method in the context of cross-dataset generalization. We then follow up our experiments with evaluating the reliance of our method on labels in comparison to full supervision. We then perform an ablation study in order to evaluate the impact of each major component in our solution, along with the choice of fusion strategy. In the end, we evaluate the impact of different augmentations as well as the impact of using negative pairs in the self-supervised training.

\begin{table}[!t]
\caption{\textcolor{black}{
Performance of our proposed method on MotionSense dataset is presented in comparison to other self-supervised and contrastive methods.}} \label{tab:other_methods_comparision_Motionsense}
\centering
\begin{tabular}{l l c c c}
\hline
\textbf{Sup.} & \textbf{Method}  & \textbf{Val.} & \textbf{F1} & \textbf{K}\\ 
\hline\hline 

{\multirow{6}{*}{\rotatebox[origin=c]{90}{FS}}} & CNN\cite{saeed2019multi}  & Scheme 1           &  0.903          & 0.876  \\ 
& CNN\cite{saeed2019multi}  & Scheme 2          & 0.924        & 0.903  \\ 
& DeepConvLSTM\cite{haresamudram2020masked}  & Scheme 2           &  0.852          & -  \\ 
& Transformer classifier\cite{haresamudram2020masked}  & Scheme 2           &  0.803          & -  \\ 
& CPC (1D Conv Encoder)\cite{haresamudram2021contrastive}   & Scheme 1            &  0.891          & -  \\ 
& Ours  & Scheme 1           & 0.911      & 0.887 \\ 

\hline
{\multirow{8}{*}{\rotatebox[origin=c]{90}{SS}}} & Multi-task\cite{saeed2019multi}  & Scheme 1           &  0.939          & 0.923  \\ 
& Multi-task\cite{saeed2019multi}    &Scheme 2           & 0.900            & 0.874  \\ 
& Masked recon.\cite{haresamudram2020masked}   & Scheme 2           & 0.880           & -  \\ 
& CPC\cite{haresamudram2021contrastive}    & Scheme 1           & 0.882            & -  \\ 
& SimCLR\cite{tang2020exploring}   & Scheme 2           & 0.942            & -  \\ 
& Motion Prediction\cite{rahimi2021self}    & Scheme 1           & 0.934            & -  \\ 
& Ours   & Scheme 1          & 0.934           & 0.916  \\ 
& Ours    & Scheme 2          & 0.943      & 0.928 \\ 
\hline

\end{tabular}
\end{table}

\begin{table}[!t]
\caption{\textcolor{black}{
Performance of our proposed method on HAPT dataset in comparison to other self-supervised and contrastive methods.}}
\label{tab:other_methods_comparision_HAPT}
\centering
\begin{tabular}{l l c c c}
\hline
\textbf{Sup.}  & \textbf{Method}  & \textbf{Val.} & \textbf{F1} & \textbf{K}\\ 
\hline\hline

{\multirow{3}{*}{\rotatebox[origin=c]{90}{FS}}}  & CNN\cite{saeed2021sense}  & Scheme 1           &  0.897          & 0.880  \\ 
 & CNN\cite{saeed2021sense}  & Scheme 2          & 0.899        & 0.883  \\ 
 & Ours   & Scheme 1           & 0.883      & 0.874 \\ 
\hline
{\multirow{5}{*}{\rotatebox[origin=c]{90}{SS}}}  & Transformation recognition \cite{saeed2021sense}  & Scheme 1                & 0.896            &  0.880 \\ 
 & Transformation recognition \cite{saeed2021sense}  & Scheme 2               & 0.899            &  0.882  \\ 
 &  Motion Prediction \cite{rahimi2021self}   & Scheme 1              & 0.900            & -  \\  
 & Ours  & Scheme 1            & 0.911           & 0.896  \\ 
 & Ours  & Scheme 2                   & 0.915          & 0.901  \\ 
 \hline
\end{tabular}
\end{table}

\begin{table}[!t]
\caption{\textcolor{black}{
Performance of our proposed method on HHAR dataset in comparison to other self-supervised and contrastive methods.}}
\label{tab:other_methods_comparision_HHAR}
\centering
\begin{tabular}{l l c c c}
\hline
\textbf{Sup.}  & \textbf{Method}  & \textbf{Val.} & \textbf{F1} & \textbf{K}\\ 
\hline\hline
{\multirow{5}{*}{\rotatebox[origin=c]{90}{FS}}} & CNN\cite{saeed2019multi}  & Scheme 1           &  0.808          & 0.778  \\ 
& CNN\cite{saeed2019multi}  & Scheme 2          & 0.728        & 0.682  \\  
& SCN\cite{saeed2020federated}   & Scheme 1              & 0.82           & 0.80 \\ 
& SCN\cite{saeed2020federated}   & Scheme 2              & 0.73            & 0.69  \\  
& Ours  & Scheme 1           & 0.803      & 0.772 \\  
\hline
{\multirow{6}{*}{\rotatebox[origin=c]{90}{SS}}} & Multi-task\cite{saeed2019multi}  & Scheme 1       &  0.804           & 0.772  \\ 
  & Multi-task\cite{saeed2019multi}  & Scheme 2          & 0.786            & 0.756  \\  
    & SCN\cite{saeed2020federated}   & Scheme 1              & 0.78            &0.76  \\  
 & SCN\cite{saeed2020federated}   & Scheme 2              & 0.82            & 0.79  \\  
& Ours & Scheme 1    & 0.826  & 0.792   \\  
& Ours  & Scheme 2    & 0.850          & 0.822  \\ 
\hline

\end{tabular}
\end{table}

\noindent \textbf{Comparison to the state-of-the-art.} We compare the performance of our solution to other self-supervised and contrastive methods in the literature evaluated on MotionSense, HAPT, and HHAR datasets (in a subject-independent manner) and present the results in Tables \ref{tab:other_methods_comparision_Motionsense}, \ref{tab:other_methods_comparision_HAPT}, and \ref{tab:other_methods_comparision_HHAR}. F1 and kappa scores presented in the tables indicate that our localized time-frequency contrastive learning approach outperforms the prior self-supervised approaches in both validation strategies (Schemes 1 and 2), setting a new state-of-the-art for the datasets. 
Moreover, it can be observed that our self-supervised model outperforms the approaches trained in a fully-supervised manner on all three datasets, indicating the capacity of our solution to learn highly generalized and discriminative representations.

\vspace{5pt}
\noindent \textbf{Cross-dataset generalization.}
Here, we evaluate the performance of our proposed approach in learning robust and general representations by applying transfer learning across different data sources. To achieve this goal, we pre-train our self-supervised network on one dataset and successive to transferring the frozen pre-trained layers, fine-tune the HAR MLP heads using the remaining three datasets. Considering the large number of participants and activities, MobiAct \cite{chatzaki2016human} dataset is used for the pre-training and the performance of the unsupervised learned features were evaluated across MotionSense, HAPT, and HHAR datasets. Tables \ref{tab:Mobiact_Motionsense}, \ref{tab:Mobiact_HAPT}, and \ref{tab:Mobiact_HHAR} compare the performance of our proposed solution to other approaches that have applied transfer learning for MotionSense, HAPT, and HHAR datasets, respectively. It is observed that our approach overtakes the other self-supervised methods in learning generalized features. Moreover, performance improvement is observed in comparison to training the network from the scratch on the same data-source (results presented in Tables \ref{tab:other_methods_comparision_Motionsense}, \ref{tab:other_methods_comparision_HAPT}, \ref{tab:other_methods_comparision_HHAR}), further validating our argument on the generalizability and transferability of the learned representations using our solution.

\begin{table}[!t]
\caption{\textcolor{black}{Results of evaluating the generalization ability of the learnt representations in cross-dataset transfer learning. The encoders are pre-trained on MobiAct dataset, and utilized for feature extraction on MotionSense dataset.}}
\label{tab:Mobiact_Motionsense}
\centering
\begin{tabular}{l c c c}
\hline
\textbf{Method}  & \textbf{Val.} & \textbf{F1} & \textbf{K}\\
\hline\hline
Multi-task\cite{saeed2019multi}        & Scheme 2           & 0.909         & 0.884  \\ 
Masked recon.\cite{haresamudram2020masked}                & Scheme 2           & 0.799             & -  \\ 

Ours        & Scheme 1                  & 0.926            & 0.906  \\ \hline

\end{tabular}
\end{table}

\begin{table}[!t]
\caption{\textcolor{black}{Results of evaluating the generalization ability of the learnt representations in cross-dataset transfer learning. The encoders are pre-trained on MobiAct dataset, and utilized for feature extraction on 
HAPT dataset.}}
\label{tab:Mobiact_HAPT}
\centering
\begin{tabular}{l c c c}
\hline
\textbf{Method}  & \textbf{Val.} & \textbf{F1} & \textbf{K}\\
\hline\hline
Transformation recognition \cite{saeed2021sense}                & Scheme 2           & 0.849            & -  \\ 
Ours        & Scheme 1                & 0.891            & 0.873  \\ \hline

\end{tabular}
\end{table}

\begin{table}[!t]
\caption{\textcolor{black}{Results of evaluating the generalization ability of the learnt representations in cross-dataset transfer learning. The encoders are pre-trained on MobiAct dataset, and utilized
for feature extraction on 
HHAR dataset.}}
\label{tab:Mobiact_HHAR}
\centering
\begin{tabular}{l c c c}
\hline
\textbf{Method}  & \textbf{Val.} & \textbf{F1} & \textbf{K}\\
\hline\hline
Multi-task\cite{saeed2019multi}                 & Scheme 2           & 0.755            & 0.713  \\ 
SCN\cite{saeed2020federated}                  & Scheme 2           & 0.75            & 0.71  \\ 
Ours        & Scheme 1        & 0.818            & 0.789  \\ \hline

\end{tabular}
\end{table}

\vspace{5pt}
\noindent \textbf{Ablation study and fusion strategy.} In this study, we designed our solution based on the hypothesis that extracting representations from both time and localized time-frequency domains of a signal improves the performance of the network in downstream tasks such as HAR. We explore this design choice by evaluating the performance of our network while using only one domain at a time. 
In order to accomplish this, we first report the results of ablated versions of our solution, i.e., $\mathcal{H}^{sca}_{HAR}$ and a $\mathcal{H}^{sig}_{HAR}$, separately.
The results of these two networks are presented and compared to the proposed fusion-based solution in Table \ref{Fusion}. It is evident that removing each branch results in a performance drop in comparison to the full solution, demonstrating the advantage of utilizing both domains. 

Next, we evaluate the choice of fusion strategy by replacing the score-level fusion used in our model with early (feature-level) fusion.
Here, we concatenate the feature maps extracted by the final convolutional layers of the signal and scalogram encoders. The fused features are then fed to an FC layer with 64 neurons, followed by another layer with the size of the activity classes. This is followed by a softmax layer. 
The results are presented in in Table \ref{Fusion}. We observe that the score-level approach used in our model, despite its simplicity, is better than the feature-level fusion. This could be due to the fact that the embeddings obtained from each learner are from domains that are different enough for their concatenation to not produce a meaningful feature set without the use of any domain-regularization.

\begin{table}[!t]
\caption{\textcolor{black}{The results of the ablation experiments to study the impact of different components of the proposed solution are presented. Furthermore, the performance of the method with two choices of fusion strategies (score-level and feature-level) are compared.}\label{Fusion}}
\centering
\begin{tabular}{l | c c | c c | c c}
\hline
\textbf{Datasets} & \multicolumn{2}{c}{\textbf{MotionSense}} & \multicolumn{2}{c}{\textbf{HAPT}}  & \multicolumn{2}{c}{\textbf{HHAR}}\\ 

\textbf{} & \textbf{F1} & \textbf{K} & \textbf{F1} & \textbf{K} & \textbf{F1} & \textbf{K}  \\ 
\hline\hline

Signal HAR     & 0.920  & 0.900  & 0.879   & 0.856   & 0.811  & 0.774          \\ 
Scalogram HAR      & 0.910  & 0.886    & 0.876   & 0.854  & 0.753  & 0.718         \\ 
Early fusion       & 0.926  & 0.906   & 0.894   & 0.876   & 0.816  & 0.785         \\ 
Late fusion        & 0.934  & 0.916   & 0.911   & 0.896   & 0.826  & 0.792        \\ 
\hline

\end{tabular}
\end{table}

\vspace{5pt}
\noindent \textbf{Impact of data augmentation.}
Here, we evaluate the impact of different temporal and time-frequency augmentations on our method. First we train the signal learner with eight different transformations individually, and utilizing the pre-trained encoder for HAR. Table \ref{tab:transformation_signal} presents the performance on MotionSense, HAPT, and HHAR datasets. We observe that on MotionSense, Time Warp produces the best results with respect to the other augmentations by a large margin. This is followed by Rotation, and Scale. On HAPT, Scale shows the best performance, followed closely by Time Warp and Rotation. Finally on HHAR, similar to MotionSense, Time Warp yields the best results, followed by Permutation and Rotation. Overall, we conclude from this experiment that Time Warp is highly effective, while Rotation and Scale are also often among the effective augmentations.

We also explore the effects of three different time-frequency augmentations on scalogram representations. As discussed earlier, given that the scalogram inputs are in the form of 2D matrices, like images, augmentations often used on images are explored. We report the results in Table \ref{tab:transformation_scalogram}, and observe that on MotionSense, Color Distortion achieves the best results, followed by Crop \& Resize, and Flip. In HAPT, however, Crop \& Resize is most effective, followed by Flip, and then Color Distortion. HHAR shows a trend identical to HAPT.

\begin{figure}[t!]
	\centering
	\includegraphics[width=0.9\columnwidth]{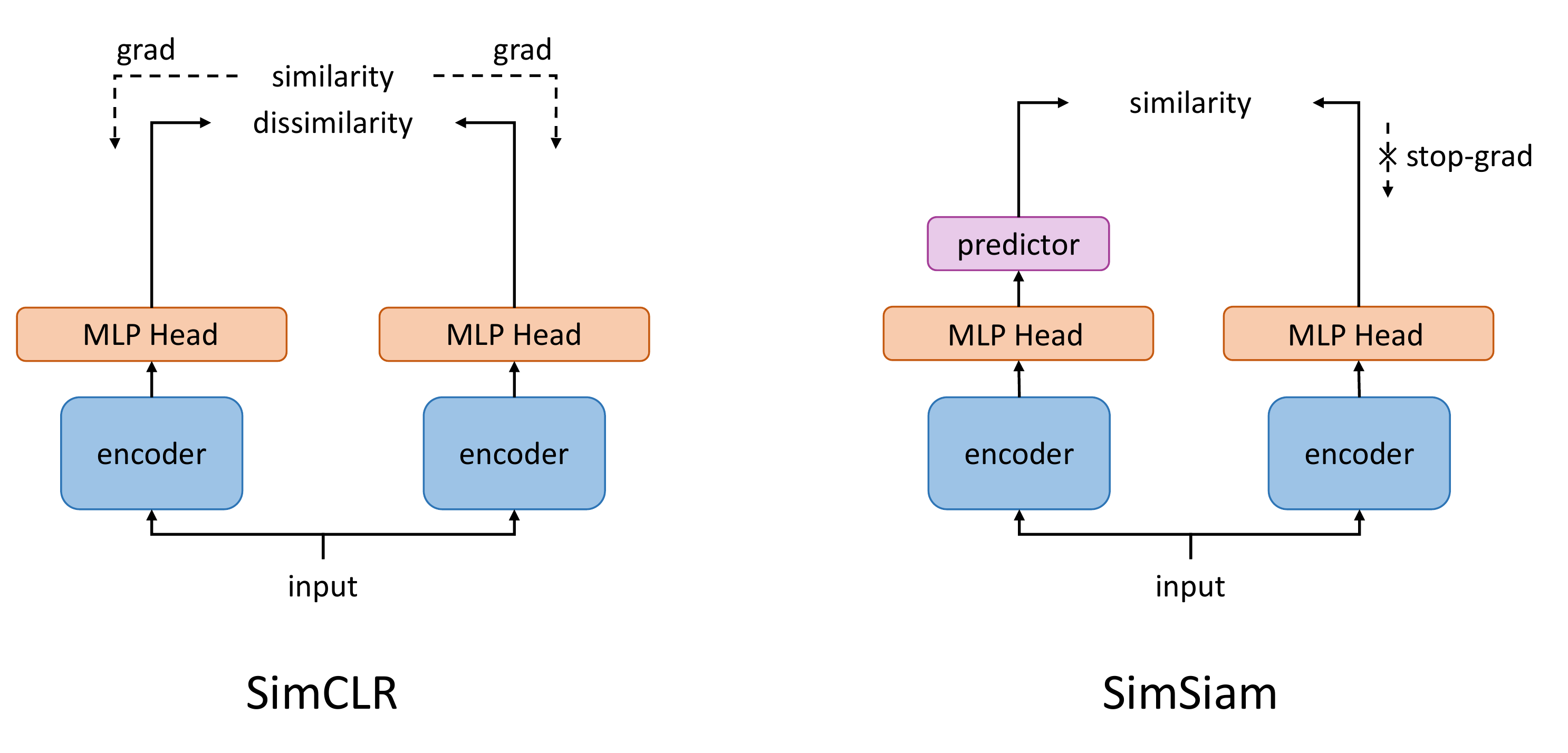}
	\caption{\textcolor{black}{A high-level comparison of SimCLR and SimSiam architectures.}}
	\label{simsiam}
\end{figure}

\vspace{5pt}
\noindent \textbf{Impact of negative pairs in contrastive training.}
While SimCLR exploits both the positive and negative pairs of data samples in a batch, it is possible to 
ignore the negative pairs and only rely on positive pairs. In fact, a number of recent self-supervised techniques have recently demonstrated strong performances according to this strategy, the most notable of which is SimSiam \cite{chen2021exploring}.
To evaluate the impact of the negative pairs, we implement our proposed solution based on SimSiam. The architecture of SimCLR and SimSiam are compared in Figure \ref{simsiam}. As demonstrated, SimSiam contains an encoder plus an MLP head \textcolor{black}{($f$)} which processes two augmented views of an input. In the next step, an MLP head called predictor \textcolor{black}{($pred$)} is applied on one branch, while a stop-gradient is applied on the other branch. The model works based on matching the output of the predictor to the other view by minimizing the negative cosine similarity \textcolor{black}{($\mathcal{D}$)} between the representation from one view and the latent representation generated by the MLP head from the other view. 
\textcolor{black}{
Here, the two augmented views are denoted as $x_1$, $x_2$. Accordingly $z_1$, $z_2$, $p_1$, and $p_2$ are defined as $z_1 = f(x_1)$, $z_2 = f(x_2)$, $p_1 = pred(z_1)$, and $p_2 = pred(z_2)$, respectively, and the loss function of the model is calculated as follows:
\begin{equation}
\mathcal{L} = \frac{1}{2}(\mathcal{D}(p_1, stopgrad(z_2)) + \mathcal{D}(p_2, stopgrad(z_1))).   
\end{equation}
}
The performance of a variant of our proposed solution based on SimSiam is presented in Table \ref{tab:simClr-simsiam}, and compared to our original solution which adopts SimCLR. As observed, the use of negative pairs through SimCLR proves advantageous as our solution outperforms the SimSiam variant on all three datasets.

\begin{table}[!t]
\caption{\textcolor{black}{Performance of the proposed method pre-trained with different signal transformations on MotionSense, HAPT, and HHAR datasets.}\label{tab:transformation_signal}}
\centering
\begin{tabular}{l | c c | c c | c c}
\hline
\textbf{Datasets} & \multicolumn{2}{c}{\textbf{MotionSense}} & \multicolumn{2}{c}{\textbf{HAPT}}  & \multicolumn{2}{c}{\textbf{HHAR}}\\ 

 & \textbf{F1} & \textbf{K} & \textbf{F1} & \textbf{K} & \textbf{F1} & \textbf{K}  \\ 
\hline\hline
Rotation        & 0.888     & 0.858     & 0.878   & 0.858    & 0.776    & 0.731   \\ 
Noise        & 0.849     & 0.811     & 0.871   & 0.852    & 0.749    & 0.700   \\ 
Scale        & 0.873     & 0.839     & 0.880   & 0.859    & 0.773    & 0.728    \\ 
Negation        & 0.845     & 0.807     & 0.875   & 0.852    & 0.771    & 0.737     \\ 
Time Flip        & 0.873     & 0.840     & 0.862   & 0.837    & 0.752    & 0.706    \\ 
Channel Shuffle        & 0.829     & 0.786     & 0.863   & 0.835    & 0.675    & 0.615   \\ 
Permutation        & 0.848     & 0.809     & 0.874   & 0.851    & 0.797    & 0.759    \\ 
Time Warp        & 0.911     & 0.889     & 0.878   & 0.854     & 0.815  & 0.779     \\ \hline
\end{tabular}
\end{table}

\begin{table}[!t]
\caption{\textcolor{black}{Performance of the proposed method pre-trained with different spatial transformations on MotionSense, HAPT, and HHAR datasets.}\label{tab:transformation_scalogram}}
\centering
\begin{tabular}{l | c c | c c | c c}
\hline
\textbf{Datasets} & \multicolumn{2}{c}{\textbf{MotionSense}} & \multicolumn{2}{c}{\textbf{HAPT}}  & \multicolumn{2}{c}{\textbf{HHAR}}\\ 

 & \textbf{F1} & \textbf{K} & \textbf{F1} & \textbf{K} & \textbf{F1} & \textbf{K}  \\ 
\hline\hline
Flip        & 0.857     & 0.818     & 0.858   & 0.835    & 0.684    & 0.637   \\ 
Crop \& Resize         & 0.866    & 0.878     & 0.867   & 0.844    & 0.741    & 0.700    \\ 
Color Distort.       & 0.900     & 0.830     & 0.851   & 0.828    & 0.668    & 0.618  \\ \hline 

\end{tabular}
\end{table}

\begin{table}[!t]
\caption{\textcolor{black}{Performance of the proposed method with two different contrastive methods on MotionSense, HAPT, and HHAR datasets.}\label{tab:simClr-simsiam}}
\centering
\begin{tabular}{l| c c | c c | c c}
\hline
\textbf{Datasets} & \multicolumn{2}{c}{\textbf{MotionSense}} & \multicolumn{2}{c}{\textbf{HAPT}}  & \multicolumn{2}{c}{\textbf{HHAR}}\\ 

 & \textbf{F1} & \textbf{K} & \textbf{F1} & \textbf{K} & \textbf{F1} & \textbf{K}  \\ 
\hline\hline
w/ negatives       & 0.934  & 0.916   & 0.911   & 0.896    & 0.826  & 0.792         \\ 
w/o negatives        & 0.908   & 0.888   & 0.908    & 0.894  & 0.743  & 0.707         \\ \hline

\end{tabular}
\end{table}

\begin{figure}[ht]
\centering
\par%
\fontsize{5pt}{6pt}\selectfont
\setlength{\tabcolsep}{0.01pt}
\begin{tabular}{cccc}
& Raw Signal & Scalogram Representations & Signal Representations\\
\makebox[0pt][r]{\makebox[10pt]{\raisebox{45pt}{\rotatebox[origin=c]{90}{MotionSense}}}} & \includegraphics[width=.3\linewidth]{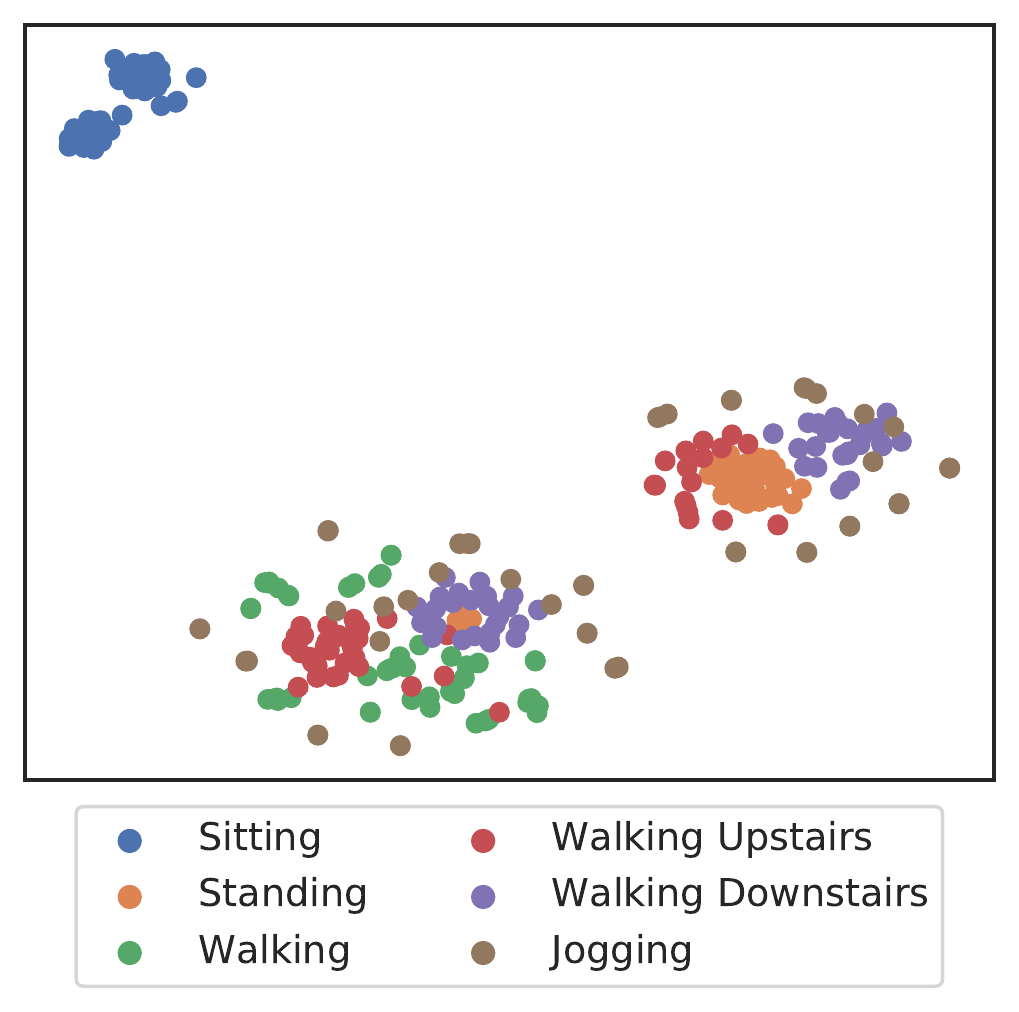} & \includegraphics[width=.3\linewidth]{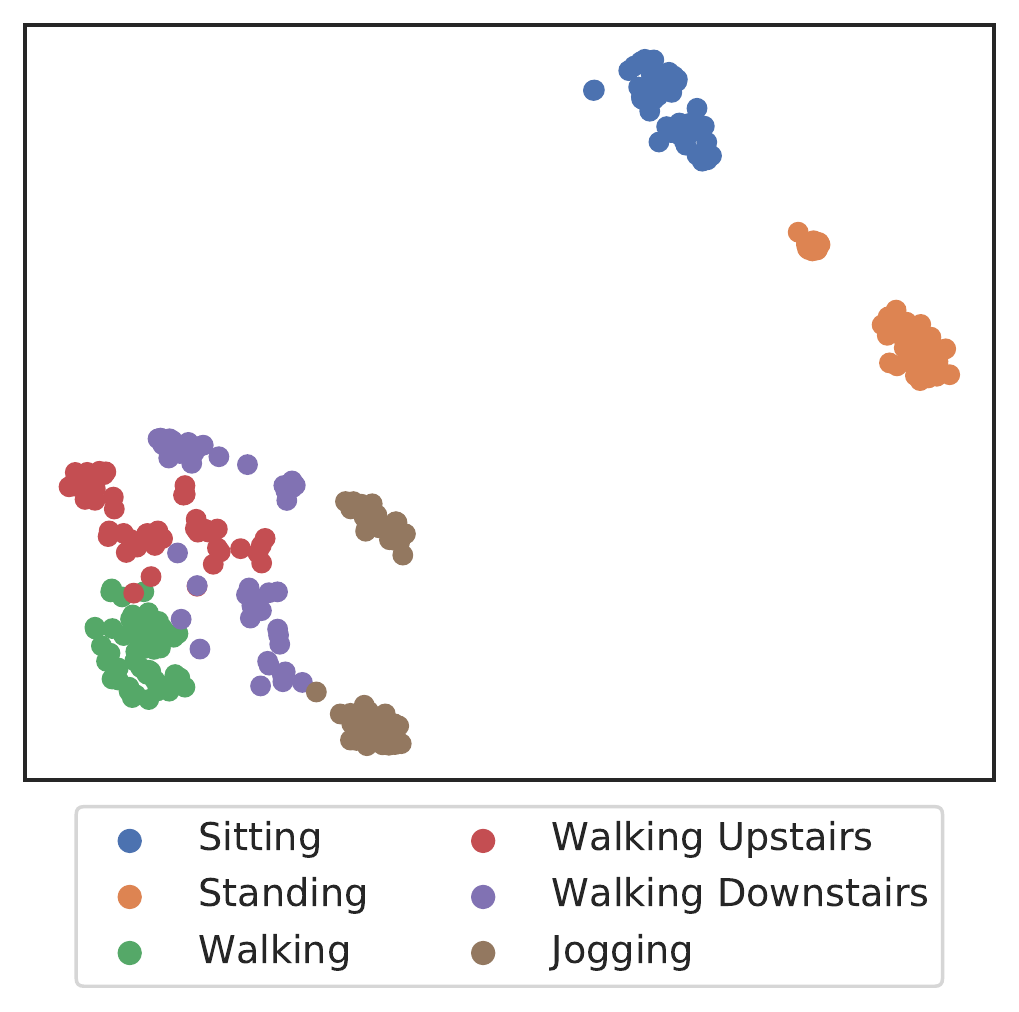} &\includegraphics[width=.3\linewidth]{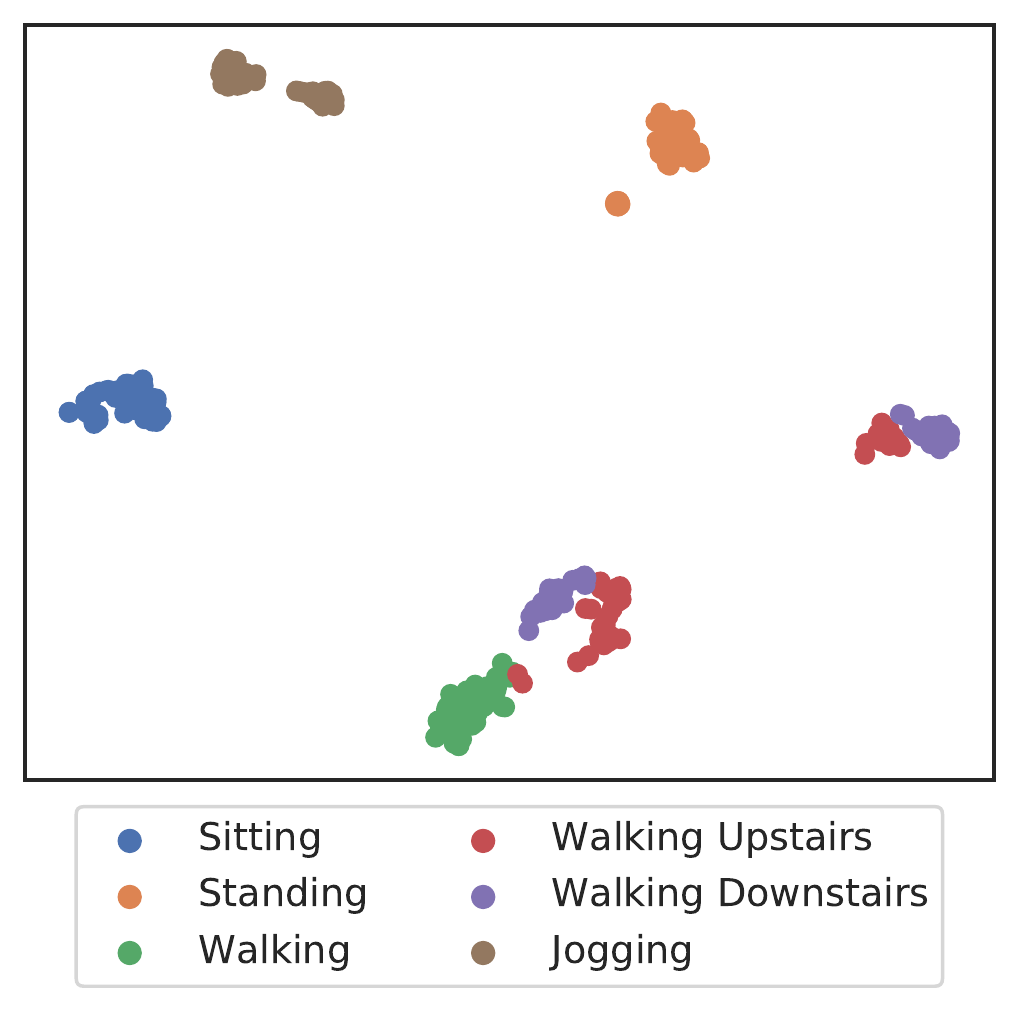}\\

\makebox[0pt][r]{\makebox[10pt]{\raisebox{45pt}{\rotatebox[origin=c]{90}{HHAR}}}} & \includegraphics[width=.3\linewidth]{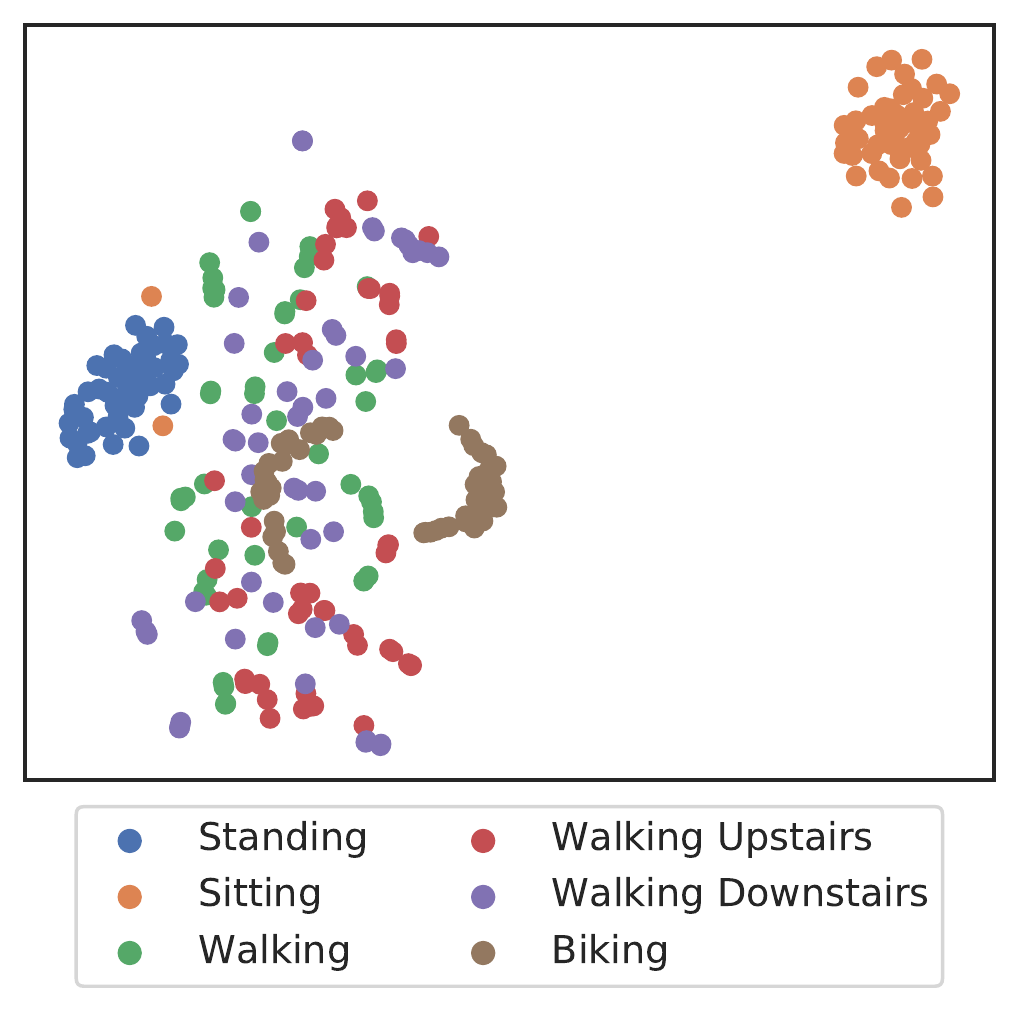} & \includegraphics[width=.3\linewidth]{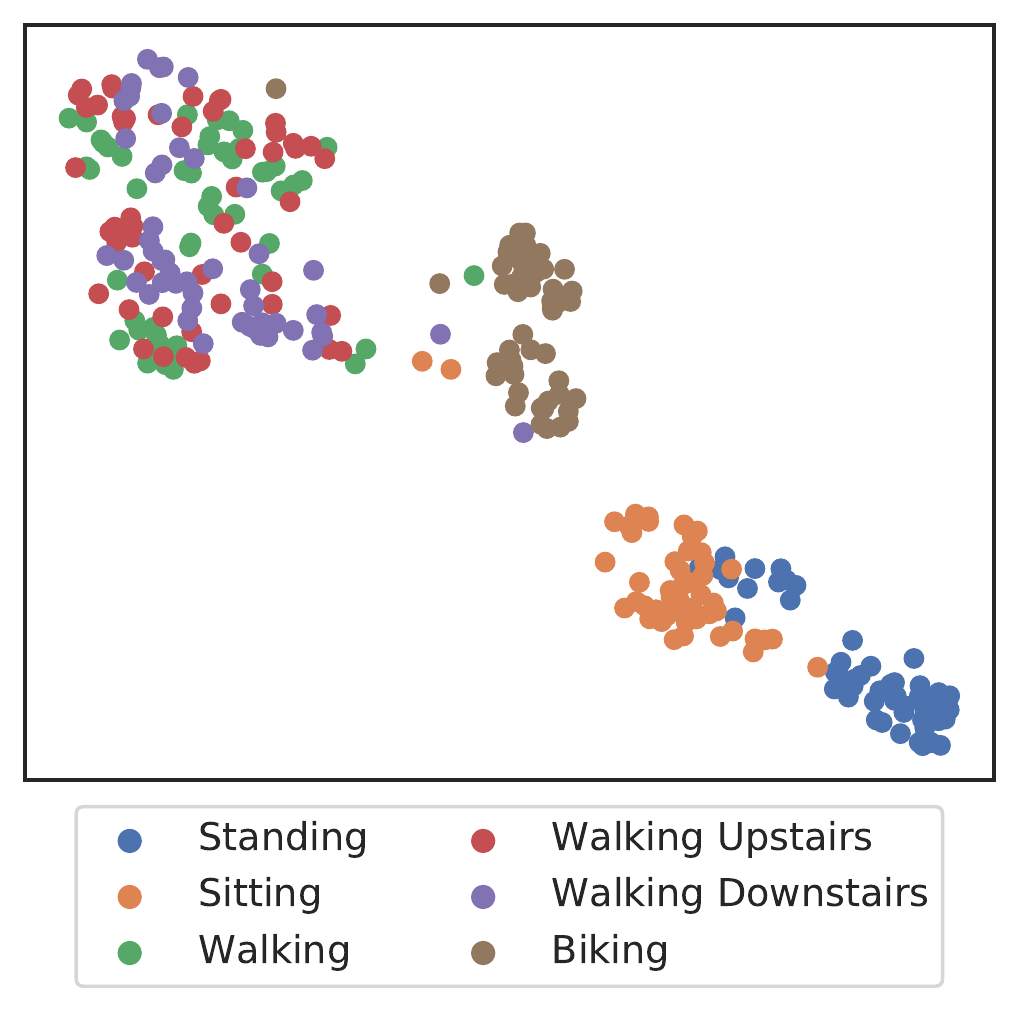} &\includegraphics[width=.3\linewidth]{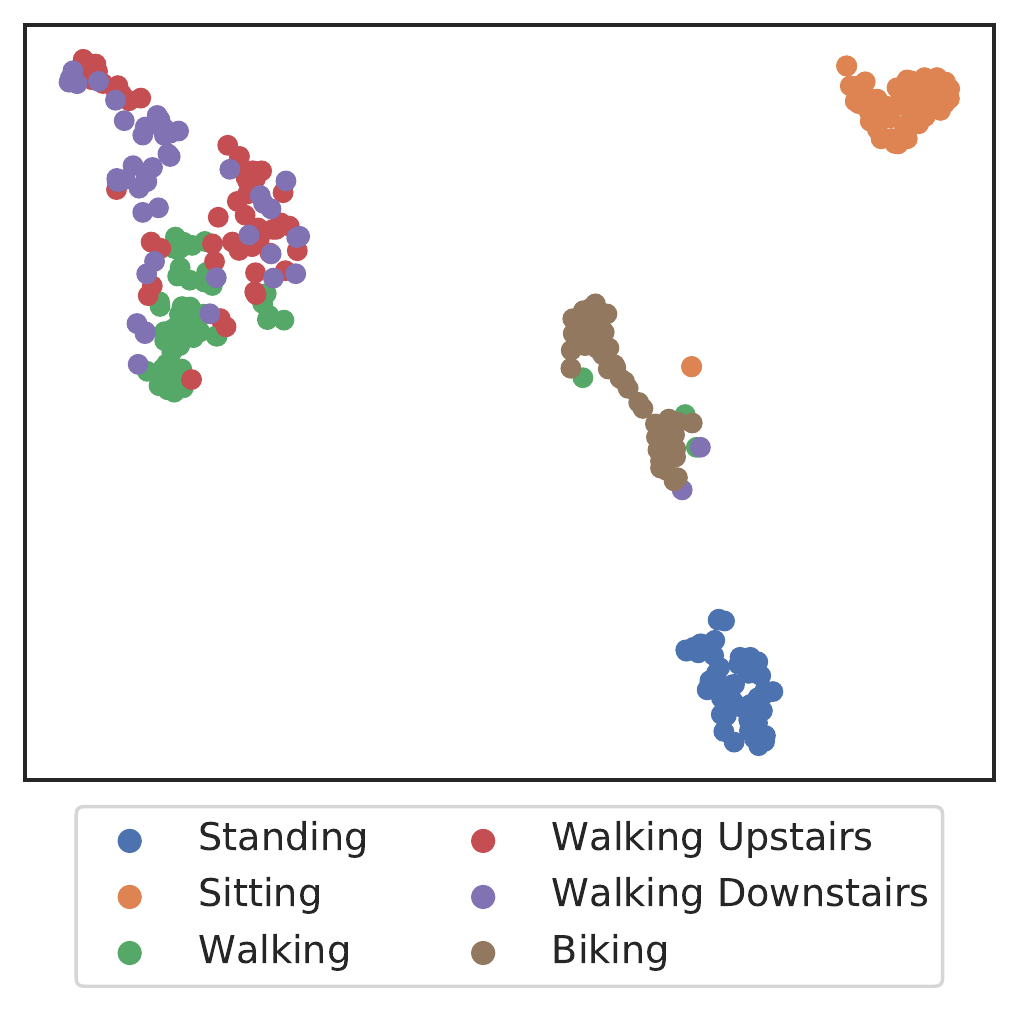}\\

\makebox[0pt][r]{\makebox[10pt]{\raisebox{58pt}{\rotatebox[origin=c]{90}{HAPT}}}} & \includegraphics[width=.3\linewidth]{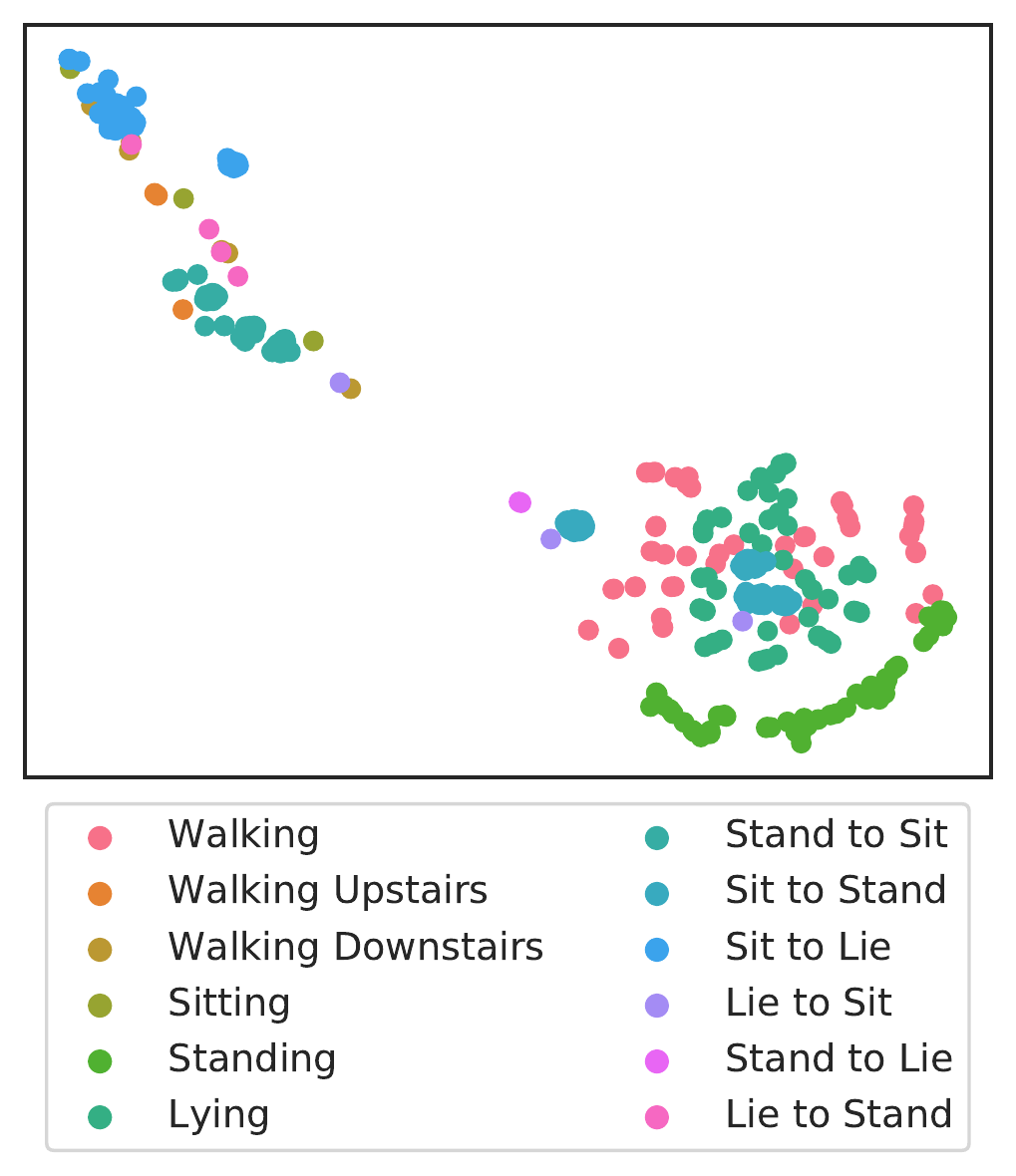} & \includegraphics[width=.3\linewidth]{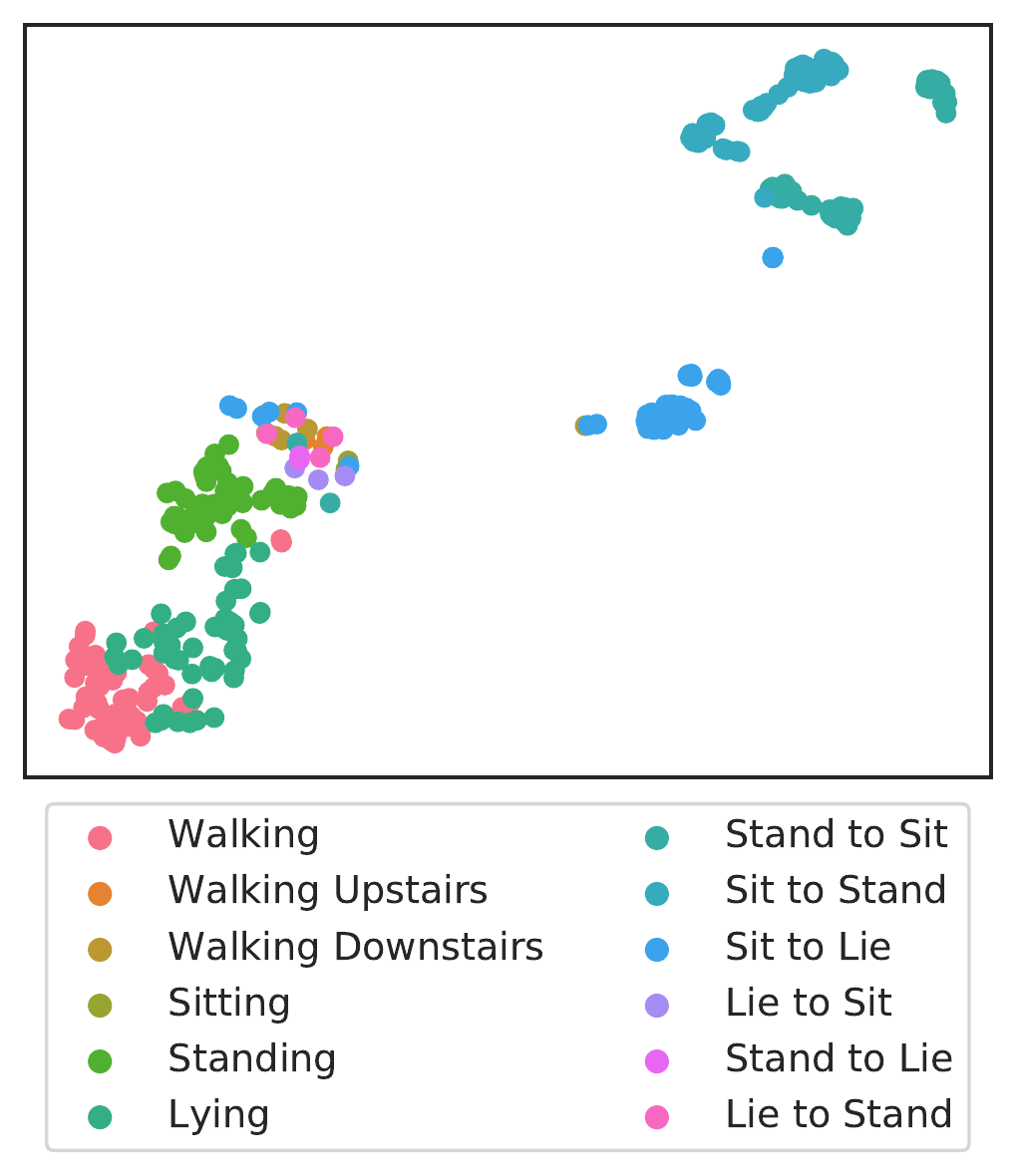} &\includegraphics[width=.3\linewidth]{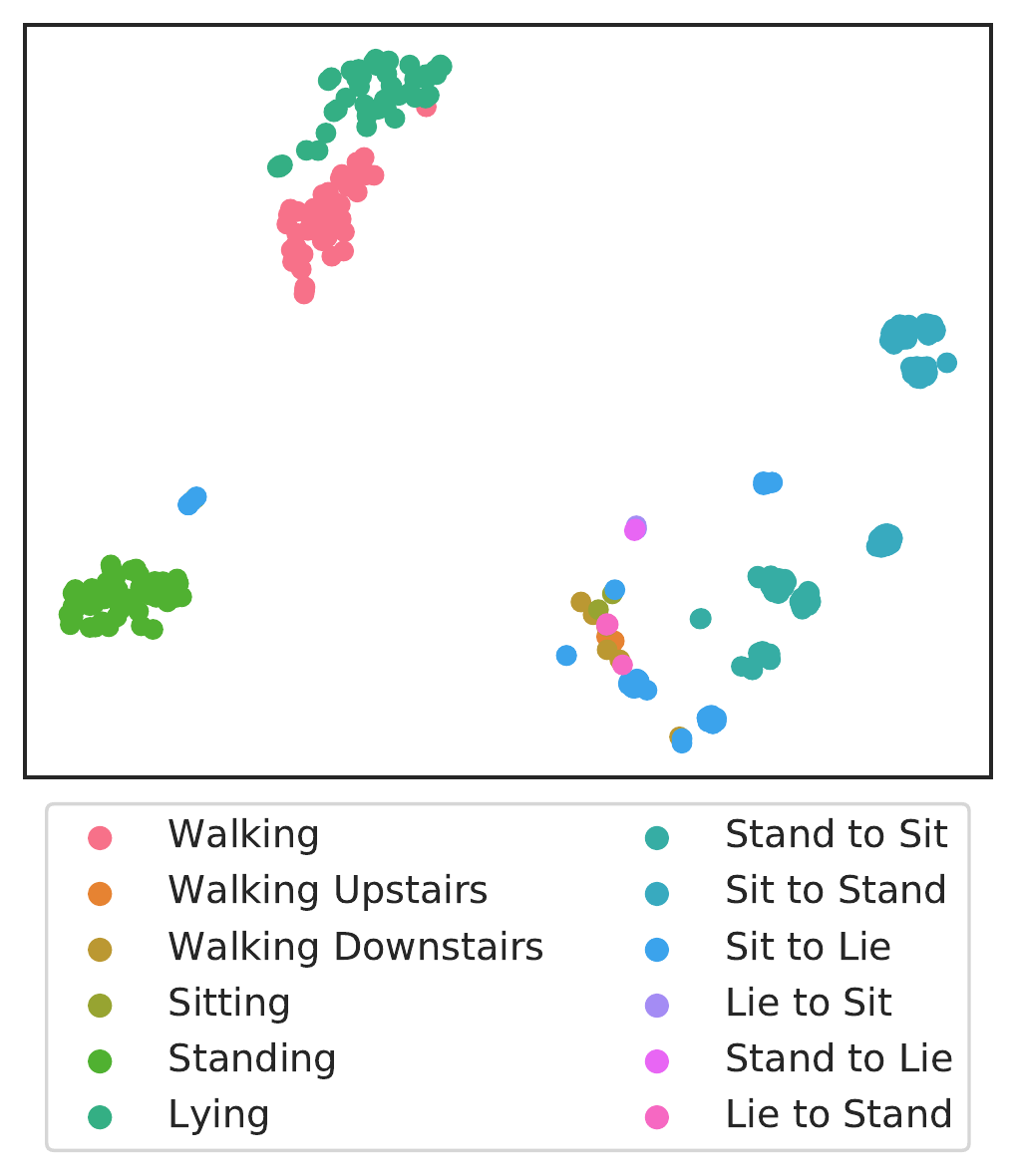}\\
\end{tabular}
\caption{\textcolor{black}{t-SNE visualization of the raw signals (left column), and learned representations by the frozen scalogram and signal encoders (middle and right columns). Each row represents the data from a different dataset, namely MotionSense, HHAR, and HAPT datasets.
    }}
\label{fig:journal_tsne}%
\end{figure}

\vspace{5pt}
\textcolor{black}{\noindent \textbf{Visualization of the learned representations.}}
In the end, we perform t-distributed stochastic neighbor embedding (t-SNE) \cite{van2008visualizing} to visualize the learned embeddings obtained using our self-supervised solution. The results are presented in Figure \ref{fig:journal_tsne}, where the first column (left) depicts the raw signals after t-SNE. Given that our method consists of two separate encoders, this analysis will yield two t-SNE plots for the learned embeddings. These are depicted in the second and third columns (middle and right, respectively), where the middle column depicts the t-SNE of the embeddings obtained from scalogram encoder while the column on the right presents the embeddings obtained from the signal encoder. Both embeddings are obtained form the frozen encoders without any fine-tuning. The three rows of the figure (top, middle, and bottom) are sample data from each of the three datasets, MotionSense, HHAR, and HAPT. In this figure, we observe that while the different activity classes are quite difficult to separate based on the raw data (left column), the two self-supervised encoders (middle and right columns) have been able to obtain embeddings that are relatively separable across activity classes. Moreover, we observe that each of the self-supervised encoders (time and scalogram) result in embeddings that are situated differently after t-SNE. This indicates the benefit of using both streams of our model as embeddings with different characteristics are extracted for different activity classes, resulting in better overall classification. This is also confirmed by our ablation study presented earlier (see Table \ref{Fusion}).

\section{Conclusion and Future Work} \label{Conclusion}
We presented a self-supervised solution for HAR which works based on contrastive learning and learns effective representations from accelerometer signals in both time and localized time-frequency domains. To tackle this goal, first, we applied a Morlet wavelet transformation on accelerometer signals to generate scalograms. Then, we implemented two contrastive networks, namely scalogram learner and signal learner which learn the data in time-frequency and time domain. Each learner consists of an augmentation component to generate different views of an input, an encoder followed by a projection head to generate two sets of representations, and a contrastive loss to maximize the similarity between the representations extracted from each sample. After the networks were pre-trained in the pretext stage, the two encoders were frozen and accompanied by FC layers to build activity classification networks. In the downstream stage, these networks were fine-tuned with labeled data, and then fused to make the ultimate decision about the activity class. We demonstrated the robustness of our solution by outperforming other self-supervised methods in HAR on MotionSense, HAPT, and HHAR datasets. Furthermore, we evaluated the ability of our solution in learning generalized representations by performing cross-datasets transfer learning. Hence, we used the MobiAct dataset for the pre-training stage, and evaluated the effectiveness of the learned representations by fine-tuning the downstream classifiers with the other three datasets. Moreover, we demonstrated the advantage of using both time and time-frequency domains by comparing the results of the fusion-based solution with its ablated versions. In the end, we demonstrated the advantage of using negative pairs by comparing our SimCLR-based solution with its SimSiam-based counterpart, observing that the use of negative pairs improves the results.

For future studies, our work could be expanded to utilize other types of neural networks for the encoders, such as RNNs for the signal encoder and the addition of residual connections for the scalogram encoder given their effectiveness in image representation learning. One of the challenges that we faced in this study, was the imbalance of activity classes, especially in HAPT dataset. In this dataset, given the high number of classes, the 6 basic activity samples noticeably outnumbered the postural transition ones. Analyzing the issue of data imbalance and how the existing imbalanced learning approaches can be effectively integrated into self-supervised representation learning, could be another direction for future studies. For example, one of the approaches used to tackle the imbalance data problem is over-sampling the minority classes. However, with data augmentation as one of the main components of contrastive-based methods, the effect of augmenting over-sampled instances may require further investigation. Lastly, in this study, we utilized a set of temporal and time-frequency augmentations to pre-train the encoders that were later used as feature extractors for activity recognition. In the future, our work could be expanded to analyze the augmentations that are more meaningful for activity signals and certain activity classes in particular. Utilizing such augmentations in the pre-training stage, could improve the performance of the encoders in representation learning, enabling them to extract features that are more activity-related.

\ifCLASSOPTIONcaptionsoff
  \newpage
\fi

\bibliographystyle{IEEEtran}
\bibliography{IEEEabrv,Bibliography}

\vfill

\end{document}